\documentclass[reprint,aps,amsmath,amssymb]{revtex4-1}

\usepackage{multirow}
\usepackage{graphicx}
\usepackage{dcolumn}
\usepackage{bm}
\usepackage{textcomp}
\usepackage{booktabs}

\usepackage[version=3]{mhchem} 

\begin{document}

\begin{abstract}
\noindent Accurate force fields are necessary for predictive molecular simulations. However, developing force fields that accurately reproduce experimental properties is challenging. Here, we present a machine learning directed, multiobjective optimization workflow for force field parameterization that evaluates millions of prospective force field parameter sets while requiring only a small fraction of them to be tested with molecular simulations. We demonstrate the generality of the approach and identify multiple low-error parameter sets for two distinct test cases: simulations of hydrofluorocarbon (HFC) vapor--liquid equilibrium (VLE) and an ammonium perchlorate (AP) crystal phase. We discuss the challenges and implications of our force field optimization workflow.
\end{abstract}

\title{Machine Learning Directed Optimization of Classical Molecular Modeling Force Fields}

\author{Bridgette J. Befort}\thanks{BJ Befort and RS DeFever contributed equally to this work.}
\author{Ryan S. DeFever}\thanks{BJ Befort and RS DeFever contributed equally to this work.}
\author{Garrett M. Tow}
\author{Alexander W. Dowling}
\author{Edward J. Maginn}\thanks{Corresponding author}
\affiliation{Department of Chemical and Biomolecular Engineering, University of Notre Dame, Notre Dame, Indiana 46556, United States}

\date{\today}

\maketitle

\section{Introduction}

Molecular modeling and simulation use computational methods to describe the behavior of matter at the atomistic or molecular level \cite{Maginn:09}. The veracity and predictive capability of molecular simulations depend critically on the accuracy of the atomic-level interaction energies, and whether the appropriate time- and length-scales are properly sampled. On one hand is a class of techniques broadly termed as {\em ab initio} or first-principles methods, where atomic interactions are determined from highly accurate quantum chemical methods \cite{Iftimie6654}. Though there are applications that necessitate these methods, {\em ab initio} energies are computationally expensive to obtain, such that quantum chemical methods are limited to relatively small systems and short timescales. On the other hand, classical molecular simulations represent the atomic interaction energies with an analytical function (a ``force field'') that can be evaluated much more rapidly than {\em ab initio} energy, enabling simulations of much larger systems and longer timescales than is possible with {\em ab initio} techniques. If force fields are highly accurate, classical molecular simulations have been shown to give accurate property predictions in several fields including protein structure refinement \cite{Heo13276}, drug discovery \cite{HOLLINGSWORTH20181129}, and energy storage \cite{Franco:2019}. 

\subsection{Developing Accurate Force Fields is Difficult}

There are two fundamentally different approaches to developing and improving force fields: bottom-up approaches, wherein parameters are calibrated so the model reproduces the results (e.g., forces, energies, and dipoles) of more expensive and accurate methods (i.e., quantum calculations) \cite{Youngs:2006}, and top-down approaches, wherein parameters are calibrated so the model matches experimental results \cite{Lobanova:2016}. Emerging bottom-up approaches use machine learning (ML) to parameterize force fields with black-box potential energy functions\cite{Noe:2020, Muller:2021:ChemRev}. Though these so-called ML force fields\cite{Parrinello:08:PRL, Muller:17:Nature} have proven successful for an increasing number of systems, the black-box nature of the potential energy function makes the models physically uninterpretable, and hinders model transferability beyond the specific training conditions. Developing accurate and transferable force fields with analytical functional forms is a difficult and laborious endeavor \cite{Harrison:2018}. Significant efforts spanning several decades have resulted in several “off-the-shelf” force fields that describe large swaths of condensed matter chemical space \cite{Rappe:92:JACS, jorgensen1996development, wang2004development, vanommeslaeghe2010charmm}. These are most commonly ``Class I'' force fields that consist of harmonic or sinusoidal intramolecular terms that describe bonded interactions, atomic partial charges that represent electrostatic interactions, and nonbonded repulsion-dispersion terms. Unfortunately, these off-the-shelf force fields can yield poor property predictions, even for relatively common compounds, particularly when they are applied in circumstances beyond the systems and conditions for which they were parameterized \cite{Martin:2006:FluidPhaseEq}. However, since they are well known and the parameter sets are widely distributed, these force fields are used in many molecular simulation studies.

For decades, force field development and optimization has been an active area of research. Several methods and tools have been developed to derive bonded intramolecular parameters and partial charges in a bottom-up fashion from quantum calculations, provided that the desired classical functional form has been selected. Common approaches include gradient-based techniques, evolutionary algorithms, or even analytical solutions \cite{resp, wang2001automatic, Prampolini.2013.PCCP, Walker.2015.JCC, Dick-Perez.2017.JCIM, wang2017building, Timerghazin:2015:JPCA, Hirao:2018:JCC}. These methods work well because the relevant quantities can be computed to a high degree of accuracy with quantum calculations, and evaluating a prospective force field parameter set is computationally trivial. However, optimizing the repulsion-dispersion parameters that are largely responsible for many macroscopic thermodynamic properties (e.g., density, enthalpy of vaporization, vapor pressure, etc.) is more challenging. Since these parameters can be difficult to derive from quantum calculations without special methods \cite{Grimme:2011}, top-down parameterization is often necessary. Yet screening thousands of prospective parameter sets is computationally expensive due to the need for sufficiently long simulations to accurately compute the relevant experimental properties. Even for relatively simple properties, a single simulation can require hours-to-days of computation time.

It is often desirable to parameterize a force field to reproduce multiple physical properties. A rigorous way to calibrate force fields with multiple properties simultaneously is to use multiobjective optimization\cite{Mostaghim:2004, Goddard:2014:JCTC, Hasse:2016:FluidPhaseEq, Vashishta.2021.SoftwareX, Vashishta:2020:CPC}, which can exacerbate the computational burden by an order of magnitude or more. In multiobjective optimization, a solution is Pareto optimal if it is not possible to improve one objective without sacrificing another objective.\cite{miettinen2012nonlinear} One approach is to weight each objective and re-solve the optimization problem for many different weights to identify Pareto optimal solutions.\cite{dowling2016framework} Thus computing a set of Pareto optimal solutions is often at least an order of magnitude more computationally expensive than single objective optimization. With much less computational effort, a finite set of candidate solutions can be classified into two groups: the non-dominated set, which comprises the solutions for which no other solution in the set offers improvement in any one objective without degrading performance in another objective, and the dominated set, comprising the solutions for which another solution offers improved performance in one or more objectives without degrading the performance in any other objective. By definition, all points in the Pareto set are non-dominated; the non-dominated set is an easy to compute approximation of the Pareto set.

Given the challenges associated with top-down optimization of the repulsion-dispersion parameters, there are fewer methods and packages available \cite{Pande:2014:JPCL, Reith:2013:Entropy} compared to intramolecular parameters and partial charge optimization. Much more frequently, attempts to improve these parameters involve {\em ad hoc} hand-tuning \cite{Murzyn:13, raabe2013molecular}, which is arbitrary and often limited to a few interaction parameters or a scaling thereof, as larger searches quickly become intractable \cite{Zhang:18}. Instead of performing multiobjective optimization, the more common approach is to use {\em ad hoc} weights to combine multiple calibration objectives into a single cost function \cite{Reith:2013:Entropy, Pande:2014:JPCL, Goddard:2014:JCTC}. However, this approach only finds a single Pareto optimal trade-off between the calibration objectives. 

\subsection{Machine Learning Directed Optimization Makes Force Field Calibration More Computationally Tractable}

The core challenges of optimizing the repulsion-dispersion parameters can be solved with a computationally inexpensive mapping between the desired physical properties and force field parameters. For certain cases, these mappings can be constructed with statistical mechanics \cite{Muller:2013:JCP, Muller:2014:AnnRev}, but this approach likely cannot be generalized to arbitrary systems. Alternatively, ML can be used to approximate the relevant mapping. For example, surrogate-assisted optimization (also known as black-box or derivative-free optimization) uses computationally inexpensive surrogate model evaluations to emulate the outputs of a complex computer simulation, e.g., computational fluid dynamics, finite element analysis, or molecular simulations. Several different types of surrogate models have been successfully applied to molecular simulations for uncertainty quantification \cite{Kino:2012:MMS, Thompson:2017:Fluids} and force field parameterization \cite{Reith:2013:Entropy, Koumoutsakos:2016:JCP, Shirts:2018:JCTC, Bauchy:2019:MRS}. Linear regression response surface models were used to predict the optimal combination of scaling factors for the charge and Lennard-Jones (LJ) parameters of General AMBER force field (GAFF) to reproduce four properties of organic liquid electrolytes. While easy to implement and moderately successful at improving the force field’s accuracy for some of the properties, this method was limited by the choice of statistically significant parameters in the response surface. \cite{zhang} For some thermodynamic properties, reweighting methods are an effective tool to test a large number of parameters without performing additional simulations \cite{Oostenbrink:2017:MolPhys, Oostenbrink:2020:JCIM, Shirts:2018:JCTC}, but care must be taken to ensure good phase space overlap between the sampled and reweighted ensembles. \cite{Shirts:2018:JCTC} Gaussian process regression (GPR) is a popular non-parametric surrogate model that smoothly interpolates between training data. Some applications of GPR in molecular simulations include ML force fields \cite{Tsuyoshi:2019:PhysSocJap, Popelier:2020:JCP, Kozinsky:2020:npj} and property prediction \cite{Duncan:2018:Microfluidics}. In Bayesian optimization, which is a special case of surrogate-assisted optimization, the uncertainty estimates from GPR (or a similar model) are directly used to balance exploration and exploitation. Recent work demonstrates Bayesian optimization can efficiently calibrate force field parameters in coarse-grained models \cite{Knapp:2019:JCIM, Yan:2020:JPCA, razi2020force}. Moreover, computationally inexpensive surrogate models can enable multiobjective optimization algorithms that go beyond {\em ad hoc} weighting \cite{miettinen2012nonlinear} to systematically explore trade-offs when calibrating multiple physical properties. 

Here, we demonstrate a new multiobjective surrogate-assisted optimization framework that uses GPRs and support vector machine (SVM) classifiers to improve existing all-atom force fields. The proposed strategy enables extremely accurate property calculations while retaining physically-motivated and interpretable functional forms. We show that the same general approach successfully optimizes force fields for two systems with very different characteristics and property objectives: hydrofluorocarbon (HFC) vapor--liquid equilibrium (VLE) and solid ammonium perchlorate (AP) crystal structure. Our results highlight the versatility of surrogate-assisted optimization approaches for top-down parameterization of all-atom force fields in a wide range of domains. The remainder of the manuscript proceeds as follows: we outline the method and provide technical details in Section \ref{sec:methods}, demonstrate the approach for the two case studies in Section \ref{sec:results}, discuss the challenges and implications of the method in Section \ref{sec:discussion}, and provide concluding remarks in Section \ref{sec:conclusions}.

\section{Methodology}
\label{sec:methods}

\begin{figure*}
\centering
\includegraphics[width=0.9\textwidth]{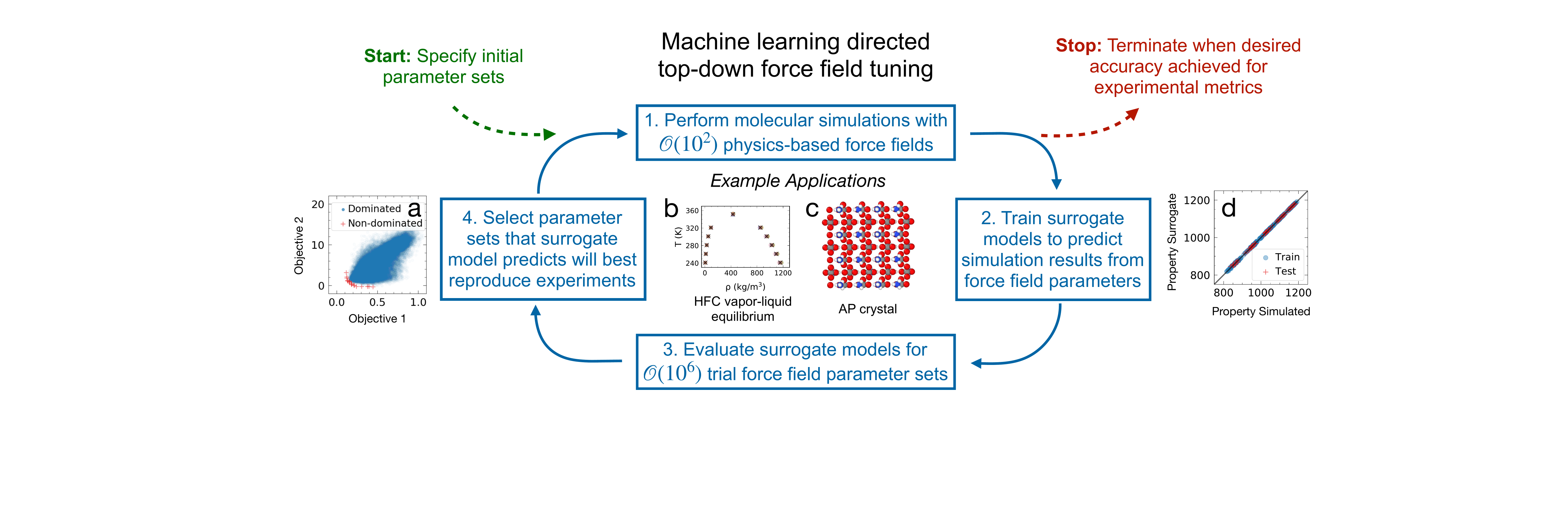}
\caption{Overview of the proposed machine learning directed force field optimization procedure. The workflow tests $\mathcal{O}(10^6)$ sets of force field parameters for every $\mathcal{O}(10^2)$ molecular simulations. The four main steps are described in the numbered boxes. Panel (a) shows the difference between dominated and non-dominated solutions for an example where the goal is to minimize two objectives. Panels (b) and (c) highlight the two example applications. Panel (d) shows an example of how the surrogate models accurately predict the outcomes of molecular simulations.}
\label{fig:overview}
\end{figure*}

\subsection{A Machine Learning Directed Force Field Optimization Workflow}
An overview of our force field optimization workflow is provided first with a more technical description given in the following subsections. Our strategy in this work is to optimize LJ repulsion-dispersion parameters, which are among the most difficult to calculate from {\em ab initio} methods \cite{Boulanger:18}. Intramolecular parameters and partial charges, which usually can be reliably and inexpensively determined from bottom-up {\em{ab initio}}-based methods, were determined from existing force fields. We stress, however, that this method can be applied to calibrate any force field parameters. 

Our force field optimization workflow is shown schematically in Figure~\ref{fig:overview}. First, domain knowledge is used to specify physically reasonable bounds on the search space for the parameters that are being optimized. Next, $\mathcal{O}(10^2)$ initial parameter sets are generated via space-filling Latin hypercube sampling (LHS). Molecular simulations are performed with each parameter set (Figure~\ref{fig:overview}, box 1), and the physical properties of interest are computed from the simulations. These results are used to train surrogate models (box 2, panel d) that predict the simulation results directly from the parameter set, and optionally, the thermodynamic state point, e.g., $T$ and $p$. Additional examples of surrogate model accuracy can be found in SI Figures~S1 and S2. The surrogate model is then used to predict the molecular simulation results for a very large number, $\mathcal{O}(10^6)$, of candidate parameter sets, once again generated with LHS (box 3). The $\mathcal{O}(10^2)$ most promising parameter sets are identified via user-selected system-specific metrics including error thresholds, separation in parameter space, and non-dominated status, from the $\mathcal{O}(10^6)$ candidate sets evaluated with the surrogate models (box 4). In multiobjective optimization, the set of non-dominated points includes all parameter sets that are not simultaneously outperformed in every dimension by any other parameter set (Figure~\ref{fig:overview}a) \cite{miettinen2012nonlinear}. Finally, the most promising parameter sets are used to initialize the next iteration of molecular simulations (box 1). The process is repeated until parameter sets are generated that provide the desired accuracy for the experimental properties of interest.

The workflow uses a combination of machine learning-based surrogate models and physics-based molecular simulations to quickly optimize force field parameters for a specific system. Physically-motivated potential energy functional forms that have proven successful over decades are retained. Whereas the molecular simulations require hours-to-days to compute experimentally measurable properties arising from a single set of force field parameters, the surrogate models can evaluate millions of parameter sets in minutes-to-hours. This means that once the surrogate models have been trained to predict the results of the molecular simulations, they enable an exhaustive search of large parameter spaces that would require $\mathcal{O}(10^7$--$10^{9})$ CPU-hours with molecular simulations. We emphasize that although the surrogate models are used to screen millions of candidate parameter sets, all of the promising candidate parameter sets are ultimately tested with physics-based molecular simulations. The role of machine learning is only to act as a surrogate for physics-based simulations, enabling the parameter search through an otherwise intractable space. The iterative procedure allows the surrogate models to improve as additional training data is collected with each iteration. The original molecular simulations are dispersed across the entire parameter space, but subsequent iterations are focused on the smaller regions of parameter space that are predicted to yield good parameter sets, enabling the surrogate models to improve in the most important regions of parameter space. The theory and technical details of each step in Figure~\ref{fig:overview} are presented in Sections \ref{sec:methods:problem} to \ref{sec:methods:step4}. Methodological details specific to the HFC and AP examples are reported in Sections \ref{sec:methods:hfcs} and \ref{sec:methods:ap}, respectively. 

\subsubsection{Problem Setup}
\label{sec:methods:problem}
The interaction potential is taken as a classical molecular mechanics force field, $U(\mathbf{r}) = f(\mathbf{r}, \boldsymbol{\zeta})$, where $U$ is the potential energy, $\mathbf{r} \in \mathbf{\Gamma}$ is the vector of position coordinates within configuration space $\mathbf{\Gamma}$, $f$ is the functional form for the potential energy, and $\boldsymbol{\zeta} = {\zeta_1, \zeta_2, ..., \zeta_N}$ are the parameters of $f$ that define the intra- and intermolecular interactions between different types of particles. Molecular simulations can be used to compute $M$ structural, thermodynamic, or dynamic properties, $\mathbf{y}^\mathrm{sim} = {y_1^\mathrm{sim}, y_2^\mathrm{sim},..., y_M^\mathrm{sim},}$ from $U(\mathbf{r})$. Depending upon the quality of $U(\mathbf{r})$, $\mathbf{y}^\mathrm{sim}$ may or may not be close to the experimental values, $\mathbf{y}^\mathrm{exp}$. The goal of this work is to refine $U(\mathbf{r})$ by optimizing $\mathcal{O}(10^1)$ force field parameters, $\boldsymbol{\zeta '} \subseteq \boldsymbol{\zeta}$, such that $\mathbf{y}^\mathrm{sim} \approx \mathbf{y}^\mathrm{exp}$ for one or more physical properties of interest. In both case studies presented here, the LJ parameters, $\sigma$ and $\varepsilon$, are optimized. Upper and lower bounds for each parameter are selected to span a wide range of physically reasonable values. The initial $\mathcal{O}(10^2)$ parameter sets are randomly selected to be space-filling within these bounds with LHS.

\subsubsection{Step 1: Perform Molecular Simulations with $\mathcal{O}$(10\textsuperscript{2}) Physics-Based Force Fields}
\label{sec:methods:step1}
Molecular simulations are performed for each parameter set with the molecular dynamics (MD) or Monte Carlo (MC) method.  For each parameter set, $\mathbf{y}^\mathrm{sim}$ is computed from the simulation output. Simulations may be performed at multiple thermodynamic conditions (e.g., $T$ and $p$) for each parameter set if the experimental data exist. Signac-flow was used to manage the setup and execution of all molecular simulations \cite{signac, signacflow}.

\subsubsection{Step 2: Train Surrogate Models to Predict Simulation Results from Force Field Parameters}
\label{sec:methods:step2}
Gaussian process (GP) surrogate models are trained to predict $\mathbf{y}^\mathrm{sim}$ as a function of the calibrated parameters $\boldsymbol{\zeta '}$. For each property, we train:

\begin{equation}
    \hat{y}_i^\mathrm{sim} = GP_i(m_i(\boldsymbol{\zeta '}), \mathrm{cov}_i(\boldsymbol{\zeta '}, \boldsymbol{\zeta'}))
\end{equation}
where $\hat{y}_i^\mathrm{sim}$ is the surrogate model prediction of $y_i^\mathrm{sim}$, $GP_i$ is the GP model for property $i$, $m_i$ is the mean function, and $\mathrm{cov}_i$ is the covariance (kernel) function. All GP models were implemented in GPFlow 2.0.0 \cite{GPflow2017}. To improve the accuracy of the GP models in regions of parameter space where $\mathbf{y}^\mathrm{sim} \approx \mathbf{y}^\mathrm{exp}$, we exclude parameter sets that result in extremely poor or unphysical results from the GP training data. We then trained SVM classifiers to predict if a parameter set was unphysical (e.g., simulation fails) so that parameter sets from these regions of parameter space could be excluded when the GP models were used to predict the results of trial parameter sets. All SVM classifiers were implemented in scikit-learn \cite{scikit-learn} with a radial basis function kernel.

\subsubsection{Step 3: Evaluate Surrogate Models for $\mathcal{O}$(10\textsuperscript{6}) Trial Force Field Parameter Sets}
\label{sec:methods:step3}
After the GP and SVM models are trained, $\mathcal{O}(10^6)$ trial parameter sets are generated with LHS. For each parameter set, the SVM and GP models are used to calculate $\mathbf{\hat{y}}^\mathrm{sim}$, the surrogate model estimates of $\mathbf{y}^\mathrm{sim}$.

\subsubsection{Step 4: Select Parameter Sets that Surrogate Models Predict Will Best Reproduce Experiments}
\label{sec:methods:step4}
Parameter sets where the surrogate models predict good agreement with experiment, $\mathbf{\hat{y}}^\mathrm{sim} \approx \mathbf{y}^\mathrm{exp}$, are selected for the next iteration. In some cases we apply an optional distance-based search algorithm (see SI Methods) to down-select only parameter sets that are far apart in parameter space.

\subsection{Hydrofluorocarbon Case Study}
\label{sec:methods:hfcs}
Force fields were independently developed for two HFCs: difluoromethane (HFC-32) and pentafluoroethane (HFC-125). Two stages of optimization were used for each HFC. The first stage used MD simulations in the $NpT$ ensemble at: 241, 261, 281, 301, and 321 K for HFC-32 and 229, 249, 269, 289, and 309 K for HFC-125. For each temperature, the pressure was set to the experimental\cite{lemmon2018nist} saturation pressure. The only property considered during the first stage was the liquid density (LD) ($\mathbf{y} = \{\rho^l\}$). In the second stage of optimization, Gibbs ensemble Monte Carlo (GEMC) was performed. The property objectives were the saturated liquid density, saturated vapor density, vapor pressure, and enthalpy of vaporization, or $\mathbf{y} = \{\rho^l_\mathrm{sat}, \rho^v_\mathrm{sat}, P_\mathrm{vap}, \Delta H_\mathrm{vap} \}$. Simulations were performed at the same temperatures used for the first stage. Four iterations of the stage 1 optimization were performed for both HFC-32 and HFC-125. Three and five iterations of stage 2 optimization were performed for HFC-32 and HFC-125, respectively.

\subsubsection{Force Field Parameters}
The functional form was taken from GAFF \cite{wang2004development}:
\begin{align}
\begin{split}
    U(\mathbf{r}) &= U^\mathrm{intra}(\mathbf{r}) + \sum_i \sum_{j>i} \frac{q_i q_j}{4\pi \epsilon_0 r_{ij}} \\ &+ \sum_i \sum_{j>i} 4 \varepsilon_{ij} \left[ \left( \frac{\sigma_{ij}}{r_{ij}}\right)^{12} - \left( \frac{\sigma_{ij}}{r_{ij}}\right)^6 \right]
\end{split}
\end{align}
where $U^\mathrm{intra}$ contains all the intramolecular terms, $r_{ij}$ is the distance between atoms $i$ and $j$, $q$ is the atomic charge, $\epsilon_0$ is the permittivity of free space, and $\sigma_{ij}$ and $\varepsilon_{ij}$ parametrize the LJ potential that describes the repulsion-dispersion interactions between atoms $i$ and $j$. The intramolecular interactions are given by:
\begin{align}
\begin{split}
    U^\mathrm{intra}(\mathbf{r}) &= \sum_\mathrm{bonds} k_r (r-r_\mathrm{0})^2 + \sum_\mathrm{angles} k_\theta (\theta-\theta_\mathrm{0})^2 \\ &+ \sum_\mathrm{dihedrals} \nu_n \left[ 1 + \cos\left( n\phi - \gamma \right) \right]
\end{split}
\end{align}
where $r_\mathrm{0}$ and $\theta_\mathrm{0}$ are the nominal bond length and angle, respectively, $k_r$, $k_\theta$, and $\nu_n$ are force constants, $n$ is the multiplicity and $\gamma$ is the nominal dihedral angle. The sums are over all bonds, angles, and dihedrals in the system, respectively. The bond, angle, and dihedral parameter for HFC-32 and HFC-125 were taken from GAFF \cite{wang2004development}. Partial charges were determined with RESP \cite{resp} as implemented in AmberTools 1.4 \cite{amber11}. The quantum electrostatic potential was computed with Gaussian 09 \cite{gaussian09} with B3LYP/6-311++g(d,p) \cite{becke1993new,stephens1994ab}. The intramolecular parameters and partial charges are reported in SI Table~S1. 

The force field optimization method was used to determine the like-interaction parameters $\sigma_{ii}$ and $\varepsilon_{ii}$ for three atom types (C, F, and H) in HFC-32 and five atom types (C1, C2, F1, F2, and H) in HFC-125. This results in 6 parameters that are optimized for HFC-32 and 10 parameters that are optimized for HFC-125. All unlike interaction parameters were computed with Lorentz--Berthelot mixing rules. For HFC-125, C1 is the carbon bonded to one carbon atom, two fluorine atoms, and one hydrogen atom, while C2 is the carbon bonded to one carbon atom and three fluorine atoms, F1 is bonded with to C1, and F2 is bonded with C2. The lower and upper bounds for each parameter were selected per-element ($\sigma$ in \AA, $\varepsilon/k_B$ in K): $3.0 \leq \sigma_\mathrm{C} \leq 4.0 $, $2.5 \leq \sigma_\mathrm{F} \leq 3.5 $, $1.7 \leq \sigma_\mathrm{H} \leq 2.7 $, $20 \leq \varepsilon_\mathrm{C}/k_B \leq 60$, $15 \leq \varepsilon_\mathrm{F}/k_B \leq 40$, $2 \leq \varepsilon_\mathrm{H}/k_B \leq 10$. The parameter bounds for each atom type in HFC-32 and HFC-125 are summarized in SI Tables~S2 and S3, respectively. 

\subsubsection{Classifier}
An SVM classifier was trained to predict parameter sets that yielded spontaneous vaporization ($\rho^l < 500$~kg/m$^3$) in MD simulations initiated at liquid density from $\boldsymbol{\zeta '}$ and $T$.

\subsubsection{GP Model}
The GP models predicted the value of a physical property from $\boldsymbol{\zeta '}$ and $T$. The LD iterations used one GP model that predicted $\rho^l$. Parameter sets with $\rho^l < 500$~kg/m$^3$ were excluded from the GP training data. The VLE iterations used one GP model for each property: $\{\rho^l_\mathrm{sat}, \rho^v_\mathrm{sat}, P_\mathrm{vap}, \Delta H_\mathrm{vap} \}$. All GP models used a radial basis function or Mat\'ern $\nu=5/2$ kernel and a linear mean function \cite{rasmussen2003gaussian}.

\subsubsection{Selecting Parameter Sets for the Next Iteration}
\label{sec:methods:hfcs:selecting}
A new LHS with 1,000,000 (HFC-32) or 500,000 (HFC-125) parameter sets was generated for each iteration. {\bf \emph{LD iterations}:} Each parameter set was evaluated with the LD SVM classifier at the highest $T$. Each parameter set was evaluated with the LD GP model at each $T$, and the root mean square error (RMSE) between the GP model prediction and experimental liquid density across all five temperatures was calculated for each parameter set. The 100 lowest RMSE parameter sets that the SVM predicted would remain liquid, and the 100 lowest RMSE parameter sets that the SVM predicted would transform to vapor, were selected for the next iteration. The low-RMSE, predicted-vapor parameter sets were included because they reflect disagreement between the SVM and GP models. After four LD iterations, parameter sets for the VLE-1 iteration were selected from the 800 simulated parameter sets. A distance-based search algorithm (see SI Methods) was used to select 25 well-separated parameter sets with RMSE $\leq$~10~kg/m$^3$. {\bf \emph{VLE iterations:}} Each parameter set from the LHS was evaluated with the LD GP model. Parameter sets predicted to yield LD RMSE $>$~25~kg/m$^3$ were discarded. This step was included to make use of the training data generated during the LD iterations since the LD GP model is very accurate after four LD iterations. The remaining parameter sets were evaluated with the four GP models trained to predict VLE properties ($\rho^l_\mathrm{sat}, \rho^v_\mathrm{sat}, P_\mathrm{vap}, \Delta H_\mathrm{vap}$). The RMSE difference between the GP model predictions and experimental values across all five temperatures was calculated for each property and parameter set. All dominated parameter sets were discarded. A parameter set is dominated if one or more parameter sets performs better than it in all of the considered objective dimensions (e.g., physical properties). The 25 parameter sets selected for the next iteration comprised the top performing parameter set for each physical property and 21 parameter sets selected from the remaining non-dominated parameter sets. A distance-based search algorithm identified parameter sets that were well-separated in parameter space.

\subsubsection{MD Simulations}
Simulations of 150 HFC molecules were performed in the $NpT$ ensemble at the experimental saturation pressure. Initial configurations were generated at 1000~kg/m$^3$. Following a steepest descent energy minimization, systems were equilibrated for 500~ps with the Bussi thermostat \cite{bussi2007canonical} and Berendsen barostat \cite{berendsen1984molecular} with $\tau_T = 0.1$~ps, $\tau_p = 0.5$~ps. The production simulations were 2.5~ns in length with the Bussi thermostat and Parrinello--Rahman barostat \cite{parrinello1981polymorphic} with $\tau_T = 0.5$~ps and  $\tau_p = 1.0$~ps. The final 2.0~ns of the production simulations were used to compute the average density.

The equations of motion were integrated with the leap-frog algorithm\cite{Hockney1989} and a time step of 1.0~fs. LJ interactions and short range electrostatics were cut off at 1.0 nm. The particle mesh Ewald method\cite{Pedersen:1995:JCP} was used to compute long-range electrostatic interactions. Analytical tail corrections to the LJ potential were applied to energy and pressure. All bonds were constrained with the P-LINCS \cite{hess2008p} method with the lincs-order and lincs-iter set to 8 and 4, respectively. Simulations were performed with GROMACS 2020 \cite{abraham2015gromacs}.

\subsubsection{MC Simulations}
GEMC simulations were performed with 1000 HFC molecules.  The initial liquid box (800 HFC molecules) was generated at the experimental liquid density and pre-equilibrated with a 5000 sweep $NpT$ MC simulation. The initial vapor box (200 HFC molecules) was randomly generated at the vapor density estimated from the ideal gas law. The combined system was simulated with GEMC. The systems were equilibrated for 10,000 MC sweeps followed by a production GEMC simulation was 90,000 MC sweeps.

LJ interactions and short range electrostatics were cut off at 1.2 nm in the liquid box and 2.5 nm in the vapor box. Long-range electrostatics were computed with an Ewald summation with a relative accuracy of 10$^{-5}$. Analytical tail corrections to the LJ interactions were applied to energy and pressure. All bonds were fixed at their nominal bond length. Simulations were performed with MoSDeF Cassandra 0.1.1 \cite{mosdefcassandra} and Cassandra 1.2.2 \cite{shah2017cassandra}.

\subsection{Ammonium Perchlorate Case Study}
\label{sec:methods:ap}
Simulations of AP were performed at 1~atm and 10, 78, and 298~K. Three properties were considered: (1) the absolute percent error (APE) from the experimental lattice parameters averaged across all three temperatures, i.e. the mean absolute percent error (MAPE), and (2) the mean of the absolute residuals of equilibrium average simulated atomic positions in reference to the experimental unit cell\cite{Choi1974} at 10~K, subsequently referred to as unit cell mean distance (UCMD), and (3) hydrogen-bonding symmetry that is present in the experimental crystal structure. Four workflow iterations were performed.

\subsubsection{Force Field Parameters}
The Class II force field of Zhu et al.\cite{Zhu2009} served as a basis for the development of a hand-tuned Class I force field. The partial charges were left unchanged \cite{Tow2018}. The Class II intramolecular bonds and angles were recast to the Class I harmonic functional forms; this process was {\it ad hoc} and involved qualitative matching to the experimental infrared spectrum. The most significant outcome of this procedure was that at 298~K the N---H stretching mode split into two separate peaks for the Class I force field, as opposed to the single peak observed by both experiment and the Class II force field. This is likely due to inherent limitations in the harmonic representation of the vibrational mode; in the context of our work, this trade-off in vibrational behavior for the simplicity and transferability of the Class I AP force field is acceptable. The LJ parameters of the hand-tuned force field were also developed with an \textit{ad hoc} approach, using similar structural metrics as described above. The hand-tuned AP force field parameters are reported in SI Table~S4.

The force field optimization workflow was applied to further optimize the $\sigma$ and $\varepsilon$ for the 4 unique atom types in the AP model, giving a total of 8 calibrated parameters. The lower and upper bounds for each parameter were as follows ($\sigma$ in \AA, $\varepsilon$ in kcal/mol): $3.5 \leq \sigma_\mathrm{Cl} \leq 4.5 $, $0.5 \leq \sigma_\mathrm{H} \leq 2.0 $, $2.5 \leq \sigma_\mathrm{N} \leq 3.8 $, $2.5 \leq \sigma_\mathrm{O} \leq 3.8 $, $0.1 \leq \varepsilon_\mathrm{Cl} \leq 0.8 $, $0.0 \leq \varepsilon_\mathrm{H} \leq 0.02 $, $0.01 \leq \varepsilon_\mathrm{N} \leq 0.2 $, $0.02 \leq \varepsilon_\mathrm{O} \leq 0.3 $. The parameter bounds are also summarized in SI Table~S5. All unlike LJ interactions were calculated with geometric mixing rules.

\subsubsection{Property Calculation Details}
In an effort to be more consistent with the refined hydrogen positions described by Choi et al.\cite{Choi1974}, the hydrogen atoms in the primitive cell were extended along their N---H vectors to match the N---H lengths that they report in Table V. To assess the symmetry that should be present in orthorhombic AP's \textit{Pnma} space group, the differences in the N---H(3)$\cdot \cdot \cdot$O(3) mirror symmetric bond lengths and angles were computed. Hydrogen bonds within 0.001~\AA~and angles within 0.3\textdegree~were considered symmetric. To determine tolerances for assessing symmetry, the manually tuned force field was utilized and the frequency of saving coordinate data over the 100~ps production run was varied between 100--10,000~fs. When saving the coordinates every 100~fs, the symmetric hydrogen bond lengths were within 0.00003~\AA~and the angles were within 0.01\textdegree~of each other. When saving the coordinates every 10,000~fs, the resolution of symmetry decreases to within 0.001~\AA~for bonds and 0.3\textdegree~for angles. For data management reasons, the coordinates were saved every 10,000~fs and the corresponding symmetry tolerances were utilized in classifying if a given parameter set was successful in reproducing the experimentally observed symmetry in the hydrogen bonding structure of AP.  

\subsubsection{Classifier}
Two SVM classifiers were trained. The first classifier predicted whether a parameter set would yield an accurate 10~K unit cell with UCMD $<$~0.8~\AA, and the second classifier predicted whether a parameter set would yield the desired hydrogen bond symmetry, as defined above. 

\subsubsection{GP Model}
Two GP surrogate models were trained. The first GP model predicted the 10~K UCMD from $\boldsymbol{\zeta '}$. Parameter sets with UCMD $\geq~$0.8~\AA~were not included in the training data. The second GP model predicted the APE of the lattice parameters from $\boldsymbol{\zeta '}$ and $T$. Both GP models used a Mat\'ern $\nu=3/2$ kernel and a linear mean function \cite{rasmussen2003gaussian}. 

\subsubsection{Selecting Parameter Sets for the Next Iteration} 1,000,000 new parameter sets were generated using LHS for each iteration. Each parameter set was evaluated with the UCMD and symmetry classifiers. Parameter sets that did not meet the UCMD threshold were discarded. The remaining parameter sets were evaluated with the two GP models. The lattice APE GP model was evaluated at $T=$ 10, 78, and 298~K for each parameter set. The mean of the lattice parameter APE at each temperature was calculated and recorded as the lattice MAPE. All parameter sets that did not meet the UCMD and lattice MAPE thresholds listed in the SI Table~S6 were discarded. When selecting parameter sets for the fourth iteration, the symmetry SVM was used to remove all parameter sets that did not meet the symmetry threshold (SI Table~S6). A total of 250 parameter sets were selected for the next iteration. All non-dominated parameter sets were selected. The remainder of the parameter sets were selected by applying an $L_1$ distance metric in scaled parameter space and the distance-based search to identify well-separated parameter sets.

\subsubsection{MD Simulations}
Simulations of orthorhombic AP were performed in the $NpT$ ensemble at 1~atm and 10, 78, and 298~K. The AP structure was taken from the 10~K data of Choi et al. \cite{Choi1974} The simulation cell comprised 378~($6\times9\times7$) unit cells. Initial velocities were drawn from a Gaussian distribution with the linear and angular momenta set to zero. A 1.0~fs time step was utilized with the time integration scheme derived by Tuckerman et al. \cite{Tuckerman2006} The equations of motions were those of Shinoda et al. \cite{Shinoda2004} Nos\'e--Hoover style algorithms were utilized for both the thermostat and barostat with relaxation times of 0.1~ps and 1.0~ps, respectively. The \textit{x}-, \textit{y}-, and \textit{z}-dimensions were allowed to fluctuate independently while maintaining an orthorhombic geometry. All simulations utilized 100~ps of equilibration followed by an additional 100~ps for generating production data. Pairwise LJ and Coulombic interactions were computed up to 1.5~nm and long-range electrostatic interactions were computed using the particle--particle particle--mesh method \cite{Hockney1989} with a relative accuracy of 10\textsuperscript{-5}. No analytical tail corrections were applied to the repulsion-dispersion interactions. All bonds were fully flexible. Simulations were performed with LAMMPS, version 7 Aug 2019 \cite{Plimpton1995}.

\section{Results}
\label{sec:results}

\subsection{Case Study: Hydrofluorocarbon Force Fields} Recent international agreements, including the 2016 Kigali Amendment to the 1987 Montreal Protocol, mandated the phaseout of high global warming potential HFC refrigerants \cite{united2006handbook}. Accurate HFC force fields that are compatible with typical all-atom functional forms are of interest as part of a broader multi-scale engineering effort to sustainably implement this phaseout. Here, we optimize force fields for HFC-32 and HFC-125, the two components of R-410a, a common household refrigerant, to accurately predict the pure-component VLE properties. While an accurate hand-tuned force field for HFC-32 exists in the literature \cite{raabe2013molecular}, the existing HFC-125 force fields are either inaccurate \cite{wang2004development} or rely on less common functional forms  \cite{fermeglia2003development,lisal1997effective,stoll2003set}, which often leads to challenges with force field transferability and simulation software compatibility. For HFC-32, we show that our strategy can develop force fields that outperform expert-created models, while for both HFC-32 and HFC-125, we demonstrate the large improvements that are possible compared against ``off-the-shelf'' models. 

\begin{figure}
\centering
\includegraphics[width=0.95\linewidth]{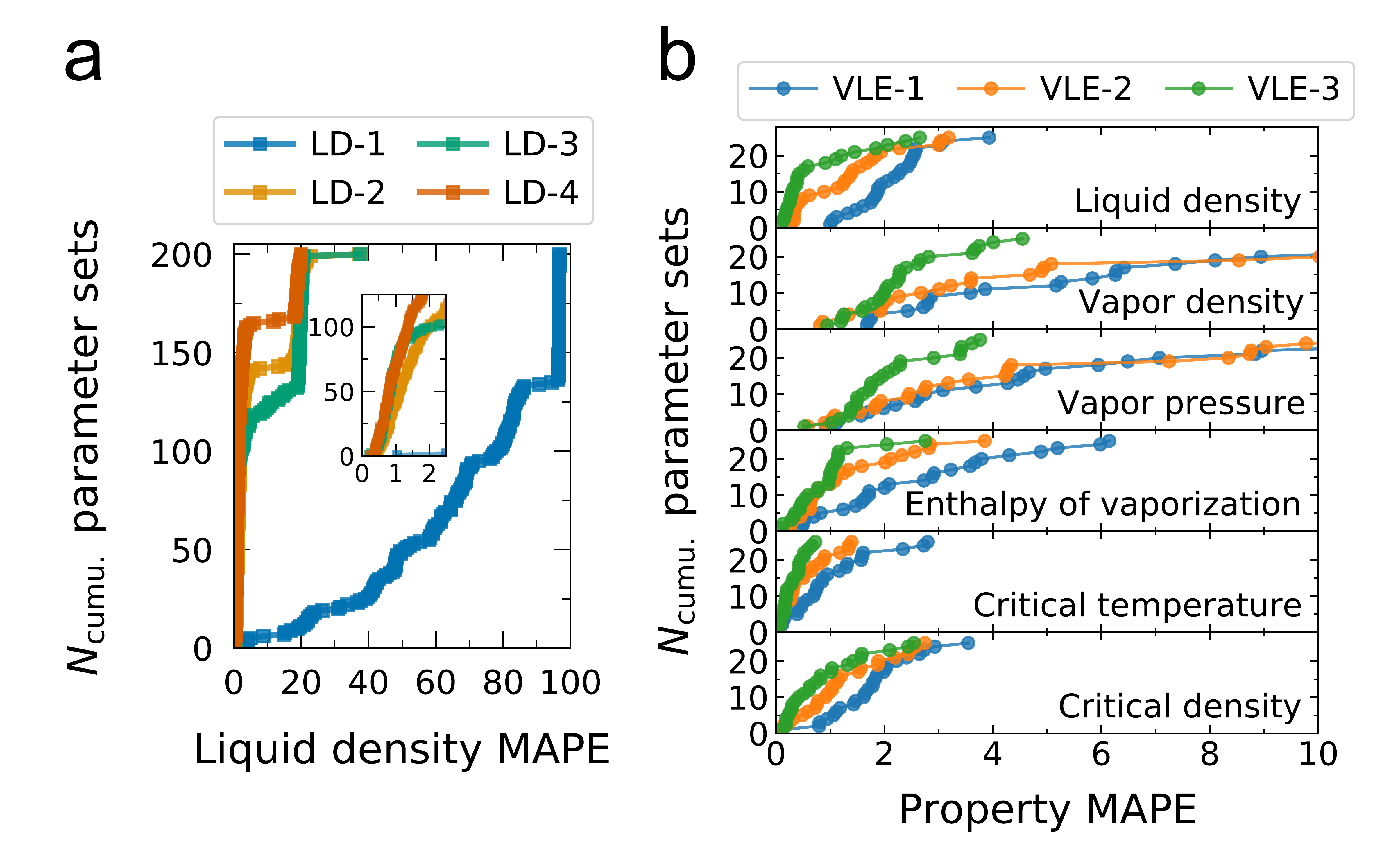}
\caption{Cumulative number of HFC-32 parameter sets generated per iteration with less than some MAPE for (a) the liquid density iterations 1--4 (LD-$n$) and (b) vapor--liquid equilibrium iterations 1--3 (VLE-$n$), where $n$ is the iteration number. Inset in panel (a) shows the LD behavior for liquid density MAPE $<2.5$\%.}
\label{fig:hfc-errors}
\end{figure}

We applied a two-stage approach to improve the HFC force fields. Our workflow was first applied to optimize the force fields to accurately predict the LD at the experimental saturation pressure for five temperatures spanning an 80 K temperature range. Following four iterations (LD-1, LD-2, LD-3, and LD-4), 25 parameter sets with low LD MAPE were used to initiate the second stage of force field optimization. In this stage, force field parameters were optimized to accurately predict VLE properties: saturated liquid density, saturated vapor density, vapor pressure, and enthalpy of vaporization. The two-stage approach has advantages: (1) the MD simulations required to compute LD in the isothermal--isobaric ensemble are computationally less expensive than the MC simulations required to compute VLE properties in the Gibbs ensemble, and (2) the stability of the Gibbs ensemble MC simulations is more sensitive to very poor force field parameters.

Figure~\ref{fig:hfc-errors}a shows the cumulative number of parameter sets that yield less than some value of the LD MAPE for each HFC-32 LD iteration. Analogous results for HFC-125 are reported in SI Figure~S3. The strength of the surrogate model approach is highlighted by the improvement from the initial liquid density iteration, LD-1, which evaluated 250 parameter sets generated directly from LHS, to the second liquid density iteration, LD-2, which evaluated parameter sets predicted by the surrogate models to yield low LD MAPE. In LD-1 fewer than 5 parameter sets had an LD MAPE below 10\%, but LD-2 yielded more than 100 parameter sets with LD MAPE below 2.5\%. Limited additional improvements are observed in LD-3 and LD-4, but additional parameter sets with low LD MAPE are nonetheless generated. Figure~\ref{fig:hfc-errors}b shows the same information for three VLE workflow iterations (VLE-1, VLE-2, and VLE-3). Consistent improvements in the saturated liquid density, saturated vapor density, vapor pressure, and enthalpy of vaporization are observed from VLE-1 to VLE-3. The results for the critical temperature and critical density also show improvement even though these properties were not explicitly included in the parameter optimization workflow. Note that the saturated liquid density in VLE-1, which evaluated 25 parameter sets generated during the LD stage, performs slightly worse than the results from LD-4 for two reasons: (1) the model vapor pressure is not precisely equal to the experimental vapor pressure, and (2) a smaller system size and shorter interaction cutoff were used to minimize the computational overhead of the LD iterations. Despite the approximation errors introduced by smaller system sizes and cutoffs, the success of our two-stage optimization strategy shows that initial iterations can be performed with less computationally expensive simulations.

\begin{figure}[ht]
\centering
\includegraphics[width=0.95\linewidth]{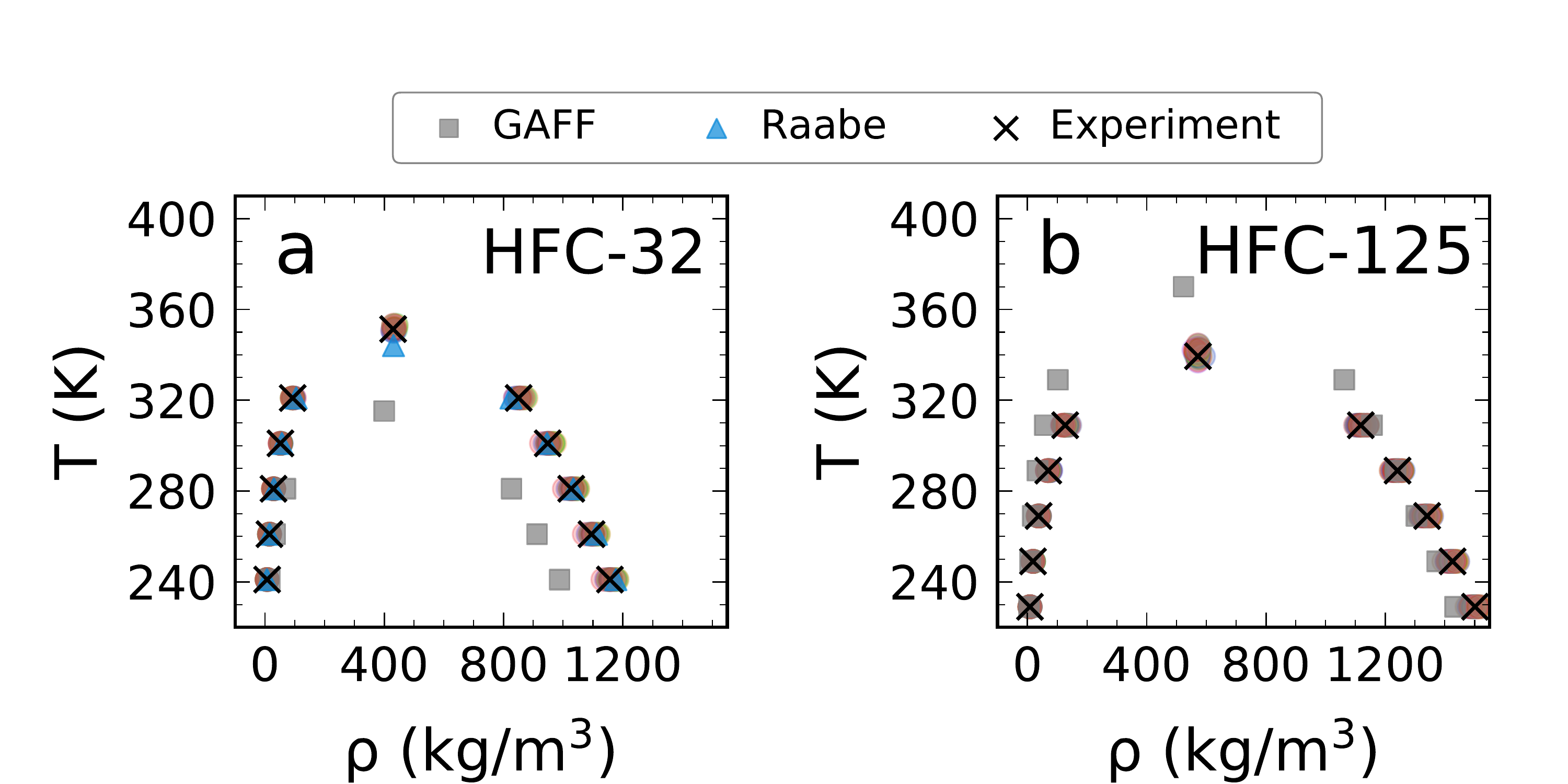}
\caption{Vapor--liquid equilibrium envelopes for (a) HFC-32 and (b) HFC-125. The 26 (HFC-32) and 45 (HFC-125) non-dominated  parameter sets identified in this work are reported as the transparent colored circles and are compared with literature \cite{wang2004development, raabe2013molecular} and experiment \cite{lemmon2018nist}. All the non-dominated parameter sets for both HFCs well reproduce the experimental values and are thus highly overlapped.}
\label{fig:vle}
\end{figure}

After completing the three(five) VLE iterations, our force field parameterization workflow yielded 26 HFC-32(45 HFC-125) non-dominated parameter sets.  Figure~\ref{fig:vle} compares vapor--liquid coexistence curves predicted by our non-dominated parameter sets with experiments \cite{lemmon2018nist} and force fields for HFC-32 and HFC-125 found in the literature. Results for vapor pressure and enthalpy of vaporization are shown in SI Figure~S4. The optimized HFC-32 and HFC-125 force fields are notably better than GAFF, and multiple optimized HFC-32 force fields give improved accuracy in all properties compared to the Raabe force field\cite{raabe2013molecular}. We chose an error threshold metric to select a subset of top-performing parameter sets from the non-dominated sets. This yielded four HFC-32 top parameter sets with MAPE of less than 1.5\% and four HFC-125 top parameter sets with MAPE of less than 2.5\% for the four properties included in the optimization workflow and the critical temperature and critical density. Comparisons of critical temperature and critical density values between experiment, the top four optimized force fields, and literature force fields for both HFCs are shown in SI Tables~S7 and S8.

\subsection{Case Study: Ammonium Perchlorate Force Field}
AP is a key ingredient in some solid rocket propellants. Experimental data for physical properties of AP are readily available and a Class II force field parameterized by Zhu et al. \cite{Zhu2009} has been used to predict \cite{Tow2018} pure AP properties at temperatures up to 298~K. The Class II functional form supplements the harmonic diagonal constants found in the more common Class I force fields through the inclusion of cross terms, namely, the stretch–stretch and stretch–bend interactions. The cross terms couple internal coordinates in an effort to better reproduce the molecular energetics as well as the dynamics of a system by accounting for anharmonic and coupling interactions. However, it is of interest to develop a Class I force field for AP to use in conjunction with existing Class I force fields for the other components of conventional solid propellant, aluminum oxide \cite{clayFF} and the polymeric binder \cite{Tow2020}. Here, we parameterize an AP force field with our force field optimization workflow; we previously had utilized hand-tuning methods to develop a Class I AP force field. We present a comparison between the conventional hand-tuning approach and our workflow. In addition to the motivation provided above, we selected solid AP as our second case study because it represents a very different system than the HFC VLE investigated in the first case study.

The properties to which we calibrated our Class I force field were: (1) UCMD at 10~K, defined as the mean of the absolute residuals of equilibrium average simulated atomic positions in reference to the experimentally observed unit cell atomic positions (low values indicate the simulation maintains the experimental AP crystal structure); (2) unit cell lattice parameter mean absolute percent error at the three temperatures of interest (10, 78, and 298~K); and (3) correct hydrogen bond symmetry. 

\begin{figure}
\centering
\includegraphics[width=0.95\linewidth]{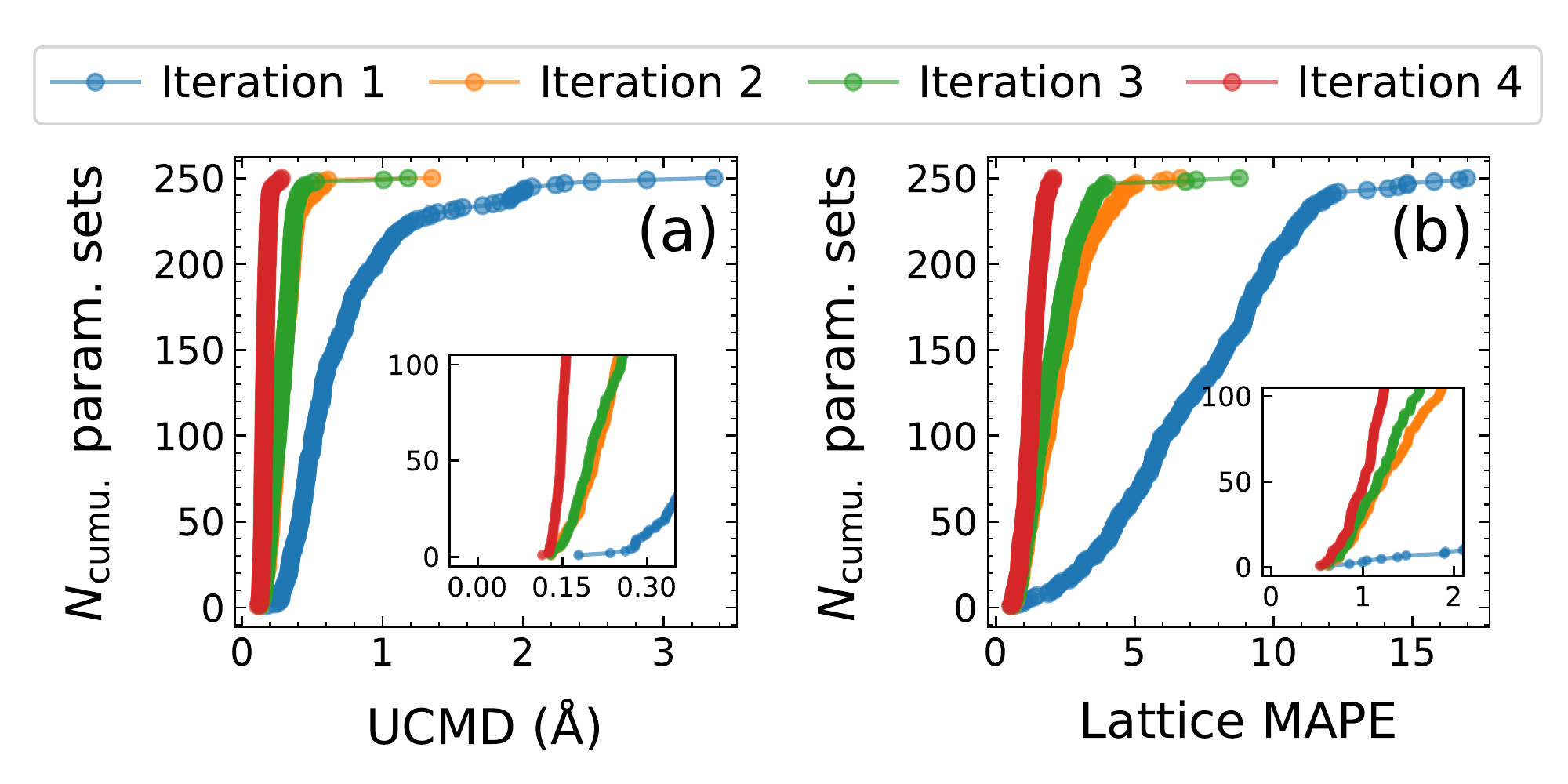}
\caption{Cumulative number of AP parameter sets per iteration with less than some value of (a) the 10 K unit cell mean distance (UCMD) and (b) the lattice MAPE. Insets have the same axis titles and focus on the improvement from iteration 3 to iteration 4. Less strict UCMD and lattice MAPE criteria were applied when selecting parameter sets for iterations 2 and 3, and stricter criteria were applied when selecting parameter sets for iteration 4. Threshold values for selecting next iteration points are shown in SI Table~S6.}
\label{fig:ap-errors}
\end{figure}
\begin{figure}
\centering
\includegraphics[width=0.95\linewidth]{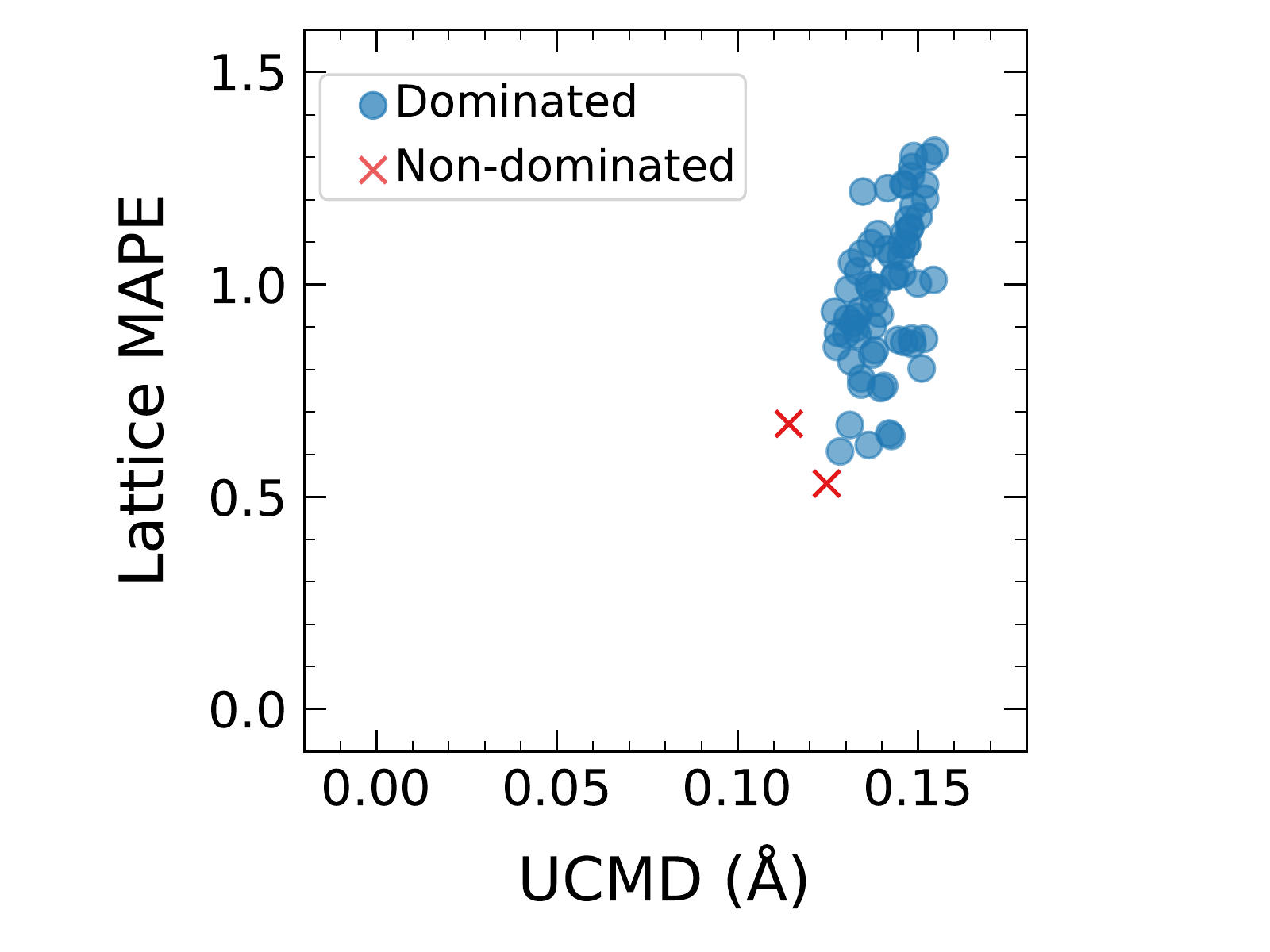}
\caption{70 AP parameter sets that yield lower UCMD and lattice parameter errors than the hand-tuned values while maintaining the correct hydrogen bonding symmetry. The red points are non-dominated and indicate our top two AP parameter sets. The blue points are dominated.}
\label{fig:ap-pareto}
\end{figure}

Four iterations of the force field optimization workflow were performed. The cumulative error plots are shown in Figure~\ref{fig:ap-errors}. Once again, we observe substantial improvement between the first and second workflow iteration. Here, the cumulative error plots also show that the criteria for selecting parameter sets for the next iteration can significantly affect the improvement in objective performance between iterations. Less strict UCMD and lattice MAPE criteria were applied when selecting parameter sets for iterations 2 and 3, and stricter criteria were applied when selecting parameter sets for iteration 4; iteration 4 showed much greater improvement over iteration 3 whereas iterations 2 and 3 are very similar. Our workflow generated 70 parameter sets over the four iterations which gave lower UCMD and lattice parameter errors than the hand-tuned values while maintaining the correct hydrogen bonding symmetry. We found two non-dominated parameter sets, as shown in Figure~\ref{fig:ap-pareto}. These two non-dominated parameter sets will subsequently be referred to as our top two AP parameter sets.  Table~\ref{tab:APTT} compares the AP results for these top parameter sets with the hand-tuned and Class II force field results.

\begin{table}
\caption{The crystal structure results for the top two AP parameter sets, ``Top A and Top B'', identified via the workflow presented in this study, the hand-tuned parameter set (HT), and the Class II parameter set of Zhu et al \cite{Zhu2009}. Lattice parameter results are reported in terms of percent error relative to experimental results \cite{Choi1974}. The UCMD results are given in \AA. \label{tab:APTT}}
\begin{center}
\begin{tabular}{lccccc}
\toprule
 Property & T (K) & Top A & Top B & HT & Class II \\
 \colrule
 \multirow{3}{4em}{Lat. \textit{a}} & 298 & -0.77 & -0.40 & -2.09 & -0.21 \\
 & 78 & -0.88 & -0.48 & -1.87 & -2.79 \\
 & 10 & -0.24 & 0.26 & -1.38 & -3.10 \\
  \colrule
 \multirow{3}{4em}{Lat. \textit{b}} & 298 & 1.13 & 0.61 & 1.96 & 7.00 \\
 & 78 & 1.11 & 0.89 & 1.68 & 8.19 \\
 & 10 & 0.63 & 0.26 & 1.16 & 8.22 \\
  \colrule
 \multirow{3}{4em}{Lat. \textit{c}} & 298 & -0.18 & -0.74 & -1.04 & 1.64 \\
 & 78 & -0.71 & -1.10 & -1.31 & 0.46 \\
 & 10 & 0.39 & 0.04 & -0.30 & 0.32 \\
  \colrule
 MAPE & --- & 0.67 & 0.53 & 1.42 & 3.55 \\
 UCMD & 10 & 0.1142 & 0.1247 & 0.1560 & 0.3485 \\
 \botrule
\end{tabular}
\end{center}
\end{table}

\clearpage
\section{Discussion}
\label{sec:discussion}

\subsection{Many Distinct Parameter Sets Yield Equally Accurate Results} 
\begin{figure}
\centering
\includegraphics[width=1.05\linewidth]{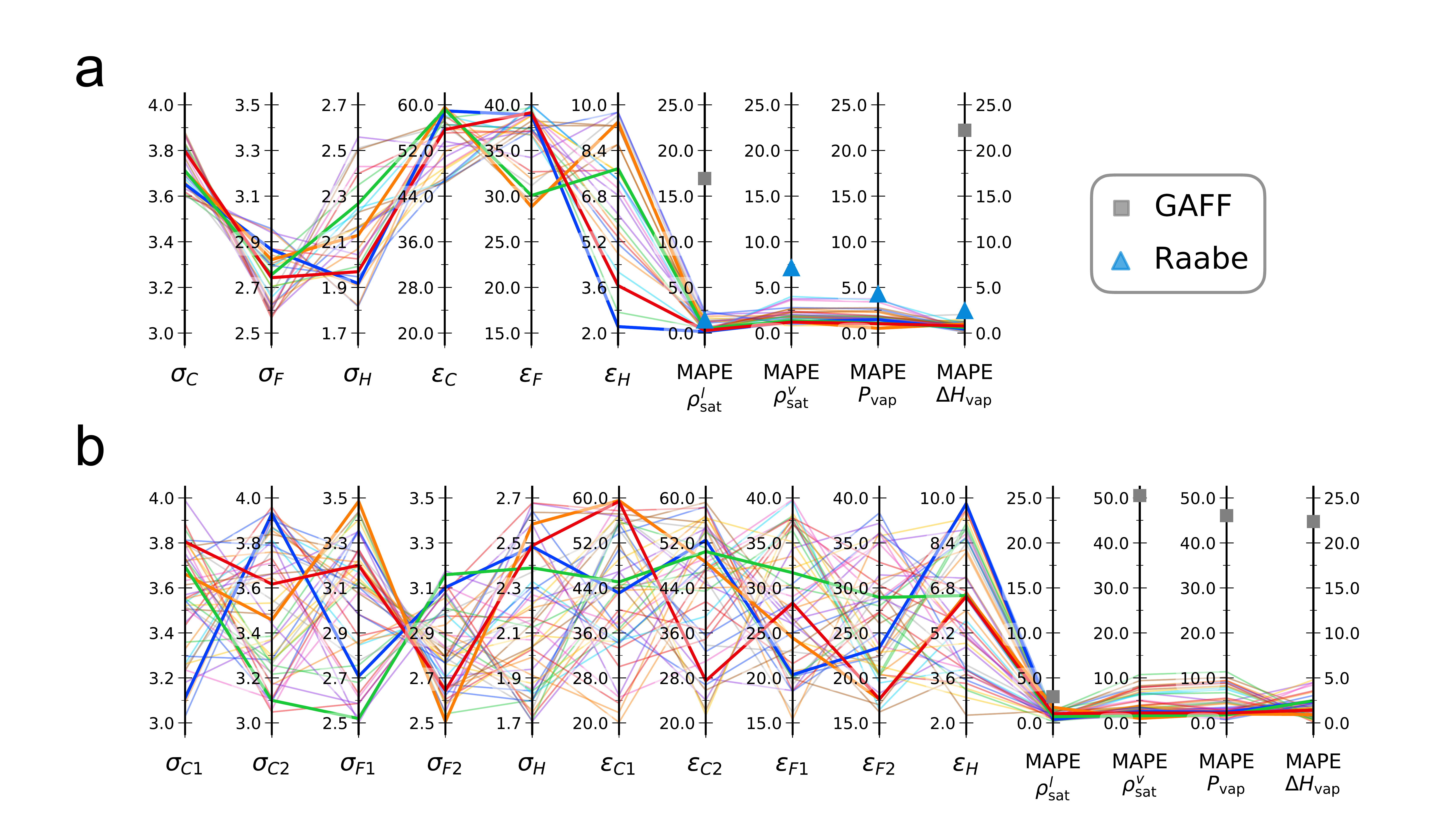}
\caption{Repulsion-dispersion parameters for (a) 26 HFC-32 and (b) 45 HFC-125  high quality parameter sets. $\sigma$ is reported in units of {\AA} and $\varepsilon$ is reported in units of K$/k_B$. Each parameter set is connected by a different color line. Thick lines indicate the top 4 parameter sets for each molecule. The y-axes are scaled to show the full range investigated for each parameter. The final four y-axes show the performance for the training objectives. The gray squares and cyan triangles show the performance of GAFF \cite{wang2004development} and the force field of Raabe \cite{raabe2013molecular}, respectively. For HFC-32 the GAFF MAPE for $\rho_\mathrm{vap}$ and $P_\mathrm{vap}$ are not shown as they are 133 and 104, respectively.}
\label{fig:hfc-params}
\end{figure}

The conventional wisdom in molecular modeling often seems to be that there is a single ``correct'' or ``best'' set of force field parameters, but this may be a misleading way to think about force field optimization. No force field is a perfect representation of the physical world. Therefore, model limitations will result in trade-offs between different objectives, and, depending on the property priorities for a specific application, lead to different optimal parameter sets \cite{stobener2014multicriteria}. However, our results clearly show that multiple parameter sets can reproduce several experimental properties with very low error. For the HFCs, our procedure yielded 26 (HFC-32) and 45 (HFC-125) non-dominated parameter sets, which are distinctly different parameterizations, all of which display good performance on our optimization objectives and the critical temperature and density. A visual representation of the non-dominated parameter sets and their performance for the optimization objectives is shown in Figure~\ref{fig:hfc-params}. For HFC-32, where there are 6 optimized force field parameters, the non-dominated parameter sets show variation of up to $\sim$0.3~\AA~in the carbon and fluorine $\sigma$ values and up to $\sim$10~K$/k_B$ in the carbon and fluorine $\varepsilon$ values. For HFC-125, there is even larger variation in the $\sigma$ and $\varepsilon$ values among the non-dominated parameter sets. We suspect this is because there are a larger number of parameters for HFC-125 (10) than for HFC-32 (6), allowing for compensating behavior between different parameters. For example, consider $\sigma_\mathrm{F1}$ and $\sigma_\mathrm{F2}$. There is a clear compensating effect: when $\sigma_\mathrm{F1}$ is larger, $\sigma_\mathrm{F2}$ is smaller, and vice-versa. On the other hand, $\sigma_\mathrm{F1}$ and $\sigma_\mathrm{F2}$ do appear to be different, as some parameterizations of $\sigma_\mathrm{F1}$ are 0.3~\AA~larger than any of the parameterizations of $\sigma_\mathrm{F2}$. 

\begin{figure}
\centering
\includegraphics[width=0.95\linewidth]{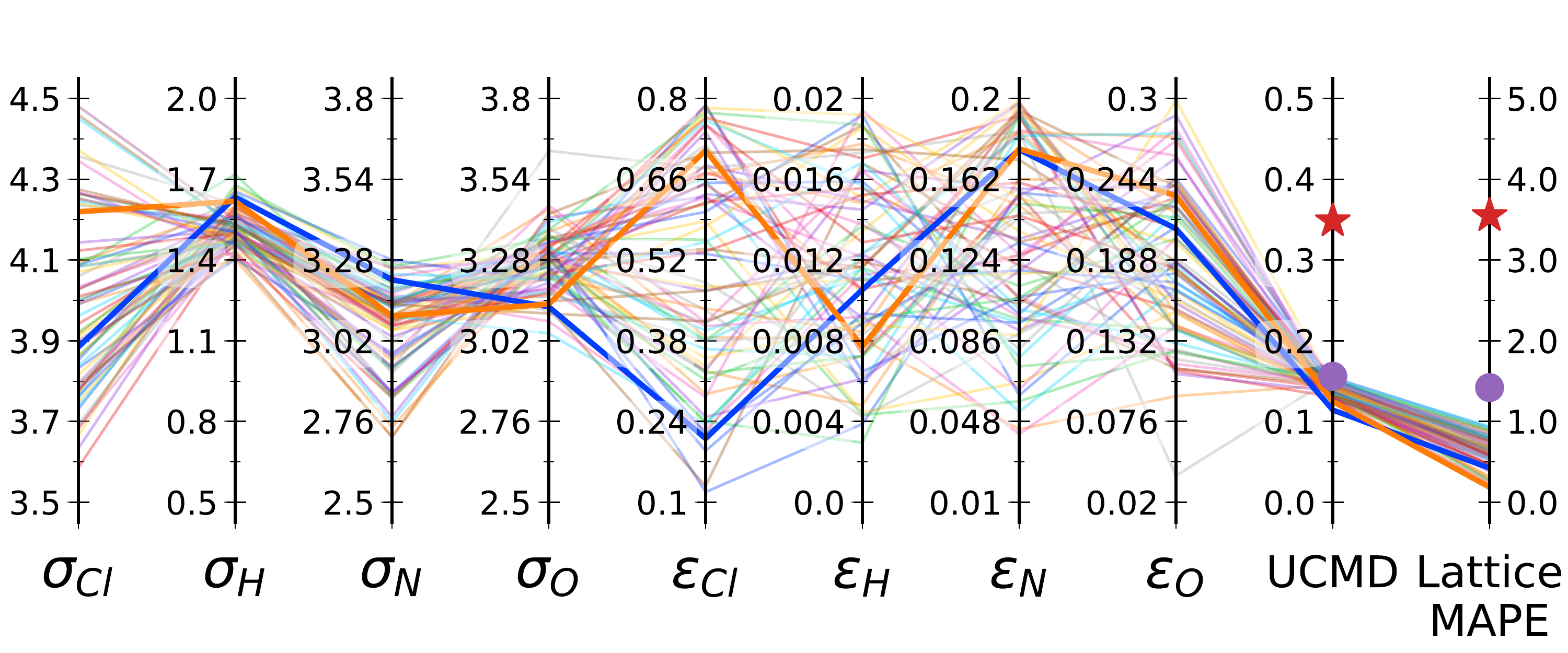}
\caption{Repulsion-dispersion parameters for the final 70 AP parameter sets. $\sigma$ is reported in units of {\AA} and $\varepsilon$ is reported in units of kcal/mol. Each parameter set is connected by a different color line. The thick lines show the top 2 AP parameter sets. The y-axes are scaled to show the full range investigated for each parameter. The final two y-axes show the training objectives. The red stars and purple circles show the performance of the Class~II force field of Zhu et al. \cite{Zhu2009} and the hand-tuned Class~I force field, respectively.}
\label{fig:ap-params}
\end{figure}

The visualizations in Figure~\ref{fig:hfc-params} suggest that the 26 (HFC-32) and 45 (HFC-125) non-dominated parameter sets are indeed distinct parameterizations, rather than closely related parameterizations with small variations along a continuous manifold of good parameters. To further investigate this question, the $L_1$ distance between the best-performing parameter set in each property and every other non-dominated parameter set was calculated and plotted against the property error (SI Figure~S5). No correlation is observed between the similarity of a parameter set to the top-performing parameter set in a given property and the property error for that parameter set. This strongly suggests that our non-dominated parameter sets are indeed distinct parameterizations. In part, this can be attributed to our procedure for advancing parameter sets to the next iteration, where we intentionally selected points that were well-separated in parameter space (Section \ref{sec:methods:hfcs:selecting}).

Similar behavior is observed in the AP system, where we identified 70 parameter sets that outperform the hand-tuned Class~I and existing Class~II force fields\cite{Zhu2009}. Figure~\ref{fig:ap-params} shows the variation in the optimized AP force field parameters. Once again, a number of distinct parameterizations yield similar accuracy for the optimization objectives. The $\sigma$ values vary by $\sim$0.3~\AA~for the hydrogen and oxygen atom types that are more exposed to intermolecular interactions, and up to as much as nearly 1.0~\AA~for the buried Cl atom type. The $\varepsilon$ values vary by as much as $\sim$0.6~kcal/mol, with the largest variation once again observed for the Cl atom type. Although there is a large variation in the individual parameter values between different parameter sets, it is the entire parameter set, taken together, that provides good performance. The results presented here do not suggest that a parameter can take any value within the ranges shown in Figure~\ref{fig:ap-params}, e.g., any value of $\sigma_\mathrm{Cl}$ between 3.5 and 4.5~\AA, and yield good performance if all other other parameter values are held constant. Rather, correlations between the different parameters enable a number of distinct yet highly accurate force field parameterizations.

Finding many distinct well-performing non-dominated parameter sets suggests the model may be overparameterized. To investigate this, we performed a local identifiability analysis by inspecting the eigenvalues of the Fisher information matrix (FIM) for the top four parameter sets for both the HFC-32 and HFC-125 models. As detailed in the SI Discussion, we find the FIM has one and five near-zero eigenvalues for HFC-32 and HFC-125, respectively, when considering only the liquid density data. This means we can only identify five (HFC-32: 6 total parameters minus 1 near-zero eigenvalue equals 5 identifiable directions, HFC-125: 10 minus 5 equals 5) parameters using only experimental liquid density data.  The corresponding eigenvectors for these near zero eigenvalues reveal the direction in parameter space in which the regression objective is flat (near zero curvature). Unfortunately, these eigenvectors do not point in the direction of a single parameter, which complicates their interpretation. More importantly, the FIM is full rank when simultaneously regressing both liquid density and VLE experimental datasets, which implies both models are locally fully identifiable. Thus, this analysis resolves one aspect of overparameterization by mathematically quantifying the importance of including multiple types of experimental data in the model calibration process. Moreover, our results suggest all of the top parameter sets are near locally optimal solutions (all with positive curvature, thus locally identifiable). 

\begin{figure}
\centering
\includegraphics[width=1.0\linewidth]{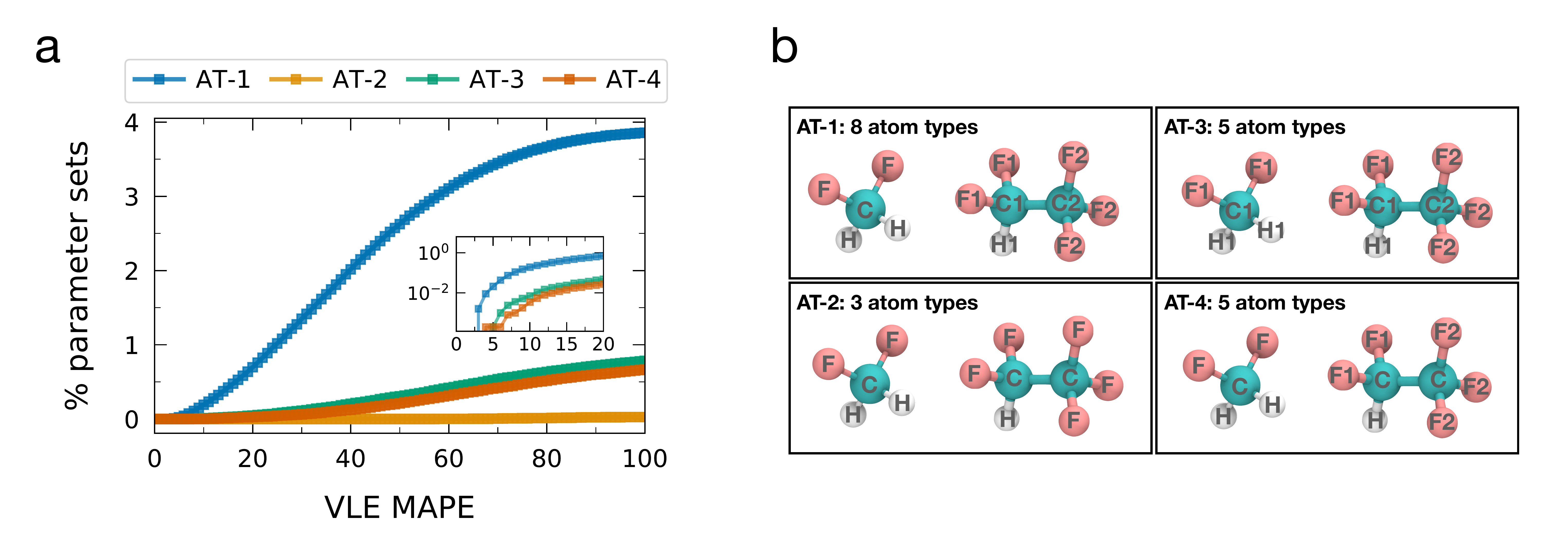}
\caption{(a) Cumulative percent of parameter sets from a large ($\mathcal{O}$(10\textsuperscript{5})) Latin hypercube that yield less than each value of MAPE for all four VLE properties. For a given MAPE, a higher percentage indicates that more parameterizations achieve at least that threshold level of accuracy. Results are shown for four different atom-typing schemes (AT-1, AT-2, AT-3, and AT-4). The inset focuses on the low-MAPE region and reports the data on a log-scale. (b) Schematic of AT-1, AT-2, AT-3, and AT-4. AT-1 is the original atom-typing scheme where no atom types were shared between HFC-32 and HFC-125.}
\label{fig:frac-viable}
\end{figure}

Another aspect of overparameterization is that we find a large number of high-quality solutions. These results are not surprising, given that many inverse problems based on engineering models have numerous locally optimal parameter sets that lead to accurate in-sample predictions.\cite{tenorio2017introduction} In this case, we hypothesize that parameterizing each molecule individually leads to many locally optimal parameter sets. Extending our method to simultaneously optimize force field parameters for an entire class of molecules (e.g., all hydrofluorocarbons) with a number of shared atom types will likely reduce the overparameterization. While we leave the development of an HFCs force field for future work, here, we explore the effects of using shared atom types for HFC-32 and HFC-125 on the number of high-quality model parameterizations. We consider four atom-typing schemes (AT-1, AT-2, AT-3, and AT-4), shown in Figure \ref{fig:frac-viable}b. AT-1 is the scheme we have used thus far; there are eight total atom types, three for HFC-32 and five for HFC-125. In AT-2, we use a total of three atom types across both molecules, C, F, and H. AT-3 and AT-4 both use five atom types, but differ in how these atom types are distributed. In AT-3, we maintain the original scheme for HFC-125, but then re-use the C1, F1, and H1 types for HFC-32. In AT-4, the C and H types are shared as they are either small or buried, while each fluorine is a different atom type. The surrogate models trained during this work were used to evaluate the performance of the different atom typing schemes. LHS was used to generate 500,000 parameter sets. First, the liquid density GP surrogate model was used to eliminate any parameter sets with RMSE greater than 100 kg/m$^3$. For each of the remaining parameter sets, the VLE GP surrogate models were used to predict the MAPE for each VLE property (saturated liquid and vapor densities, vapor pressure, and enthalpy of vaporization). Figure \ref{fig:frac-viable}a reports the percentage of the original 500,000 parameter sets that yield less than a given MAPE threshold for all four VLE properties, simultaneously. The atom-typing schemes with a reduced number of atom types have a much smaller percentage of parameter space containing low-error parameter sets. In fact, AT-2, with only 3 atom types, does not result in any parameterizations that are predicted to have below 46\% MAPE for all four VLE properties. AT-3 and AT-4 show that even with the same number of atom types, one atom-typing scheme may result in superior performance. This naturally raises another question: given different atom-typing schemes, which should be used? Recent work \cite{madin2021bayesian} demonstrates the promise of using Bayes factors to compare models with different levels of complexity (e.g., different atom-typing schemes) and make a justified selection.

Since the prior analysis was performed entirely with the predictions of the GP surrogate models, we performed molecular simulations with two top-performing parameter sets for each of the shared atom-typing schemes (AT-2, AT-3, and AT-4) in order to compute the simulated MAPE values and compare them with the surrogate model predictions. The results are reported in SI Table~S9. Overall, the surrogate model predictions were excellent, often showing less than 0.5\% MAPE deviation from the simulated MAPE. GEMC simulations for AT-2 were unstable at the highest temperature, confirming the surrogate models' prediction that AT-2 would not yield any good parameter sets. We also explored HFC-125-only force fields with a reduced number of atom types (SI Table~S10), and found that we were able to identify parameter sets with less than 3\% MAPE using only 3 atom types (C, F, and H). However, as noted above, when we attempted to use three atom types (C, F, and H) for both HFC-32 and HFC-125, no good force fields were identified. This finding is strong evidence that the fluorine atom types in HFC-32 and HFC-125 should be different (e.g., AT-4), and shows how developing parameterizations for an entire class of molecules will reduce the number of viable parameter sets.

Adding additional objective properties is a complementary strategy to reduce the number of viable parameter sets. In that case, it is important that the additional properties are orthogonal in the sense that good performance for one property is not highly correlated with good performance for another property. If property performance is highly correlated, then adding additional properties to the optimization workflow may not substantially reduce the number of viable parameter sets. The apparent overparameterization observed in this work emphasizes why tuning force fields for specific systems and using a few objective properties via relatively simple methods such as epsilon-scaling, manipulating mixing rules, or varying a single parameter value are often quite successful. However, our findings suggest that the force fields developed via these methods are most likely only one of a large number of possible parameterizations that would yield at least equal accuracy.

A further question involves how final parameter sets should ultimately be selected, given that many high-quality parameter sets are available. Our workflow is explicitly not designed to identify a single optimal set of force field parameters. Instead, it searches for and identifies high quality parameter sets with respect to all of the optimization objectives, e.g., points in the non-dominated set. Selecting a single specific parameter set from the optimized parameter sets identified by the workflow requires additional {\it post hoc} criteria that are application specific. Here, we chose non-dominated status and error thresholds for all properties. Alternative strategies include creating a weighted sum of errors in the properties based upon the desired application and domain knowledge, ranking force fields by their error in the various properties studied via statistical tests \cite{Yan:2020:JPCA}, evaluating the force field's performance for properties not included in the optimization procedure, or selecting parameter sets based upon a measure of compatibility with the force fields being used for other components of a system. One could also consider chemical intuition when selecting the final parameter sets, e.g., for HFC-125, perhaps a parameter set with more similar values for both fluorine atoms would be preferred. Though our preference is to minimize the number of {\it ad hoc} choices, ultimately, selecting the final force field for a given application will be system and application dependent and rely heavily on domain expertise.

\subsection{Maintaining a Physically-Motivated Analytical Functional Form Aids Transferability to Properties Not Included as Optimization Objectives}

One important question is whether the force field parameters developed with this workflow will yield accurate property predictions for properties not included in the optimization workflow. We have already shown that the HFC force fields developed during the VLE tuning stage result in accurate critical temperature and density even though these properties were not optimization objectives. However, these critical properties are largely determined by accurately capturing the temperature dependence of the saturated liquid and saturated vapor density, both of which were optimization objectives. To further investigate the transferability of force field parameters developed with our workflow to properties not included as optimization objectives, we examine the performance of the 25 parameter sets used during the VLE-1 iteration. These parameter sets were used for VLE-1 because they were identified as good at predicting the temperature dependence of the liquid density during the LD iterations. Figure~\ref{fig:hfc-errors} shows that when applied for VLE-1, many perform quite well for VLE properties. In fact, three of the HFC-32 parameter sets used for the VLE-1 iteration had less than 2\% MAPE in all six properties. Furthermore, when compared with GAFF, all 25 parameter sets selected from the LD stage yield better performance for all six properties. This is strong evidence that our force field optimization workflow can, with the correct optimization objectives, yield force fields that accurately predict properties beyond the optimization objectives.

The transferability of the LD-optimized parameters to VLE gives credence to our overall force field optimization philosophy, which maintains traditional analytical functional forms and uses machine learning as a guide to identify optimal parameters. However, {\it a priori}, it is unclear that there should be such a strong correlation between the liquid density and VLE properties. For many systems, accurately predicting the liquid density is a necessary, but often quite insufficient, condition for an accurate force field. We hypothesize there is a key factor that contributes to the transferability of the parameters developed during the LD iterations to VLE: the LD simulations were performed at the saturated vapor pressure across an 80 K temperature range, up to within 30 K of the experimental critical temperature. Accurately capturing the liquid density at saturation across a relatively large temperature range and avoiding spontaneous vaporization, especially at conditions closer to the critical point, requires capturing a careful balance of the cohesive energy and molecular size, which are closely related to the LJ repulsion-dispersion parameters that were calibrated. If the correlation between LD-optimized parameters and VLE properties proves applicable to other classes of molecules, it may offer a rapid method for developing force fields with accurate VLE properties.

\subsection{Selecting Good Properties for Force Field Optimization is Challenging}
When optimizing force fields for the HFC case study, we were interested in developing force fields that accurately predict HFC VLE behavior. As such, we chose to calibrate parameters to the saturated liquid and vapor densities, vapor pressure, and enthalpy of vaporization. However, these properties are expensive to compute in molecular simulations, making it difficult to evaluate a large parameter space. Therefore we used less computationally expensive LD iterations to generate good parameter sets for VLE and narrow the parameter search space. Furthermore, we continued to use the highly accurate LD GP surrogate models to screen out poor parameter sets during the VLE iterations. The success of this approach demonstrates that a cheaper ``screening'' property can be used to narrow the parameter search space drastically when good parameter sets for the screening property are a superset of the good parameter sets for the final properties of interest.

The AP case study had different challenges. The MD simulations required to predict the AP properties were computationally inexpensive, so there was no need to first use a screening property. However, it was not immediately clear what experimental properties we should target. Our first implementation attempted to reproduce the temperature dependence of the crystal lattice parameters alone; this proved ineffective, and naive in hindsight, as we generated many force fields that yielded the correct crystal lattice parameters but incorrect crystal structures. To overcome this issue, we added the 10~K UCMD as an objective because it is a measure of how accurately the force field reproduces the experimental crystal structure at 10~K. The lattice MAPE was still included to capture the temperature dependence of the crystal dimensions since the experimental unit cell coordinates are only reported at 10~K. 

The UCMD surrogate model has a notable difference from the others; whereas the other surrogate models predict a property (e.g., lattice $a$ or $p_{vap}$), the UCMD is itself an objective function. The UCMD surrogate model predicts the mean distance of all of the unit cell atoms from their respective coordinates in the experimental unit cell. By definition, this distance is zero if the simulated structure perfectly matches experiment. There are benefits to using physical experimentally measured properties compared to an objective function within the optimization workflow, including providing a clear mapping between a surrogate model and the objective metric. However, using surrogate models to predict the value of an objective function provides the opportunity to combine multiple pieces of information into a single quantity, as is the case with UCMD, which combines the distance of 40 atoms from their positions in the experimental unit cell into a single value. This strategy can drastically reduce the number of required surrogate models. In general, our experience with the AP case study emphasizes that careful thought must be given as to which experimental properties are best to target and how these should be accounted for within the workflow. Roughly 75\% of our effort for the AP case study was dedicated to identifying the appropriate experimental properties to target.

\subsection{Systematic Parameter Search Provides Insights into Model Limitations}

\begin{figure}
\centering
\includegraphics[width=0.95\linewidth]{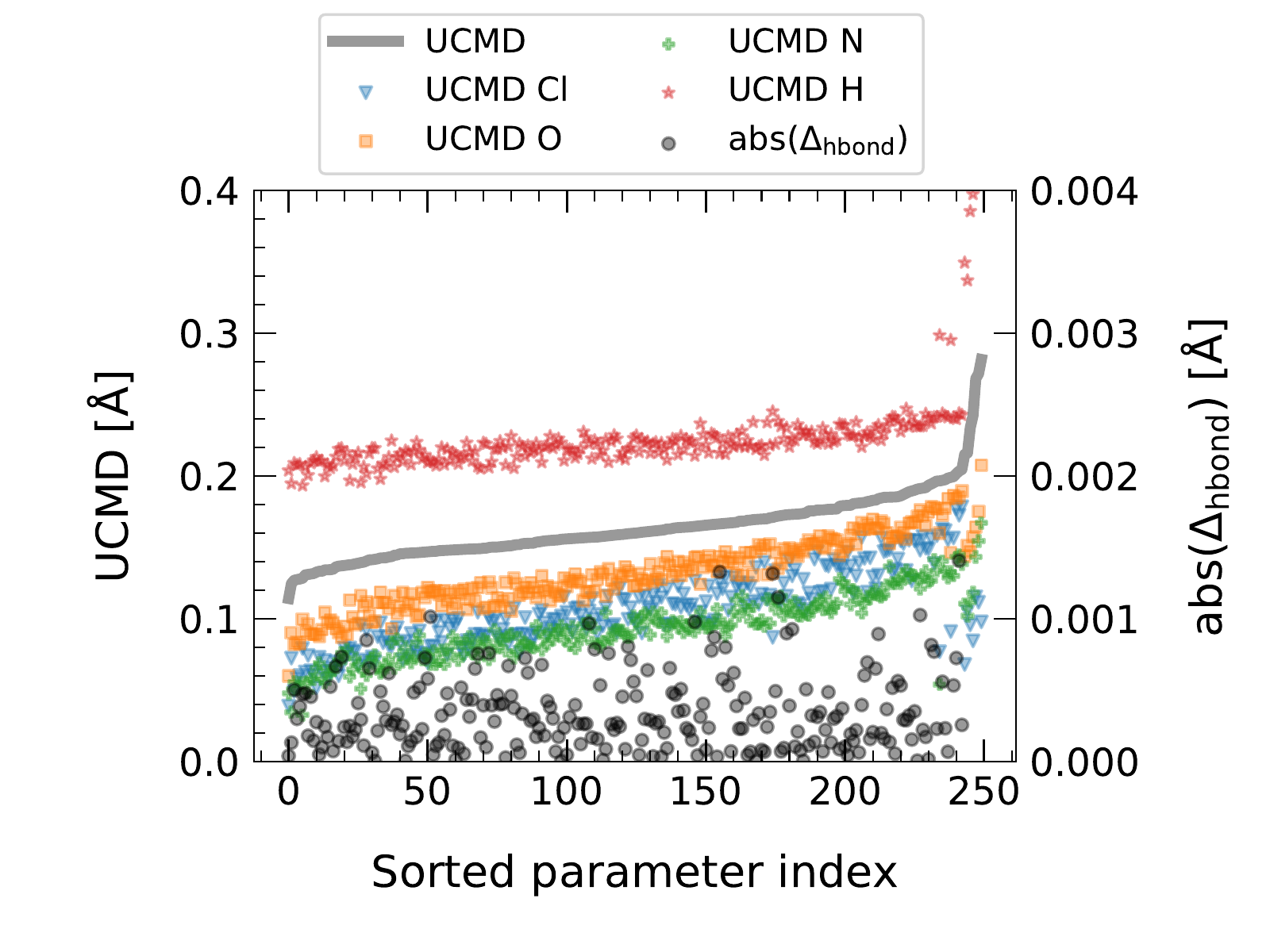}
\caption{Overall (gray line) unit cell mean distance (UCMD) compared with UCMD of the four different atom types (points) for the parameter sets tested during iteration 4 of the ammonium perchlorate force field optimization. The hydrogen bond symmetry is reported as $\mathrm{abs}( \Delta_\mathrm{hbond})$, where $\Delta_\mathrm{hbond}$ is the difference in the symmetric hydrogen bond lengths.}
\label{fig:ap-tradeoffs}
\end{figure}

The exhaustive search of parameter space enabled by our workflow provides opportunities to distinguish between inaccurate results from poor parameter sets and physical limits from our choice in force field functional form and unoptimized parameters. For example, although our workflow finds high-quality AP parameter sets, we encountered limitations that likely arise from parameters that were not calibrated, and possibly even the force field functional form that we selected. No parameter set predicted an overall UCMD of less than 0.1~\AA. Given the exhaustive search enabled by our force field optimization workflow, this suggests that there are no parameter sets capable of yielding a crystal structure with UCMD below 0.1~\AA, given the selected functional form, intramolecular parameters, and partial charges. Figure~\ref{fig:ap-tradeoffs} shows the per-element UCMD distances after iteration 4. Although the UCMD for the chlorine, oxygen, and nitrogen atoms fall between 0.1~\AA~and 0.15~\AA~for many parameter sets, the hydrogen UCMD rarely falls below 0.2~\AA. Further investigation suggests that this effect is because the N---H bond stretching is insufficiently susceptible to the three unique local hydrogen-bonding chemical environments; experiments report\cite{Choi1974} that the N---H bond lengths range between 1.028--1.058~\AA~whereas in simulations the N---H bond lengths typically cover a much smaller range --- between 1.025--1.033~\AA~--- for parameter sets that well reproduce the experimental physical properties. The N---H stretching force constant was not included in our parameterization process. However, even if it was, it is not clear that it would be possible to capture the correct bond stretching behavior and match the vibrational spectra and the N---H bond lengths with a Class I functional form. The exhaustive search provides confidence that the limitations of the model arise from the functional form and unoptimized parameters, rather than the selected parameterization.

\section{Conclusions}
\label{sec:conclusions}

We have presented a machine learning directed workflow for top-down optimization of force field parameters. By harnessing surrogate-assisted optimization, our workflow drastically reduces the number of simulations necessary to find optimal force field parameters by replacing them with computationally tractable surrogate model evaluations. We synthesize GPR and SVM surrogate models and multiobjective optimization into a generic approach to optimize all-atom force fields for realistic systems. We have applied our workflow to optimize HFC force fields to VLE properties and an AP force field to the experimental crystal structure. These case studies show that our workflow can be used for systematic exhaustive screening of parameter space and that surrogate models are highly effective at predicting both simulated physical properties and objective metrics, enabling us to find multiple low-error force fields. The approach presented here could be further combined with gradient-based methods or other approaches such as trust region surrogate-based optimization \cite{conn2000trust} to further refine the final force fields.

Based upon the success of our approach for the two disparate case studies presented here, we believe that this workflow can be applied to most molecular systems and optimization objectives, provided sufficient reference data. Surrogate models could be used to predict difficult-to-compute thermodynamic properties such as solubilities and binding energies, and transport properties such as self-diffusivity and thermal conductivity. While we have focused on calibrating repulsion-dispersion parameters in this work, this workflow could be used to calibrate any parameters within the force field in a fully top-down approach or as part of a bottom-up force field development workflow, by including {\em ab initio} data in the fitting procedure \cite{Bauchy:2019:MRS}. Additionally, we discussed the reasons for successes and limitations of the workflow, the potential challenges of applying this workflow to a particular system (i.e. choosing optimization objectives), and the questions about molecular modeling these results present. We highlight that this workflow is built on a foundation of domain knowledge in selecting the parameters to calibrate, the parameter bounds, and the experimental properties to ensure results are reasonable. 

Finally, while we believe that our workflow will enable more efficient force field development and optimization in the future, reducing the need for laborious hand-tuning practices, quantifying the workflow's efficiency was beyond the scope of this work. We can, however, anecdotally note for the AP case study that the hand-tuning approach utilized $\sim$15,000 simulations and only found 1 optimal parameter set. This is in contrast to our presented workflow, which evaluated $\sim$3,000,000 parameter sets using surrogate models, $O(10^3)$ times as many as the hand-tuning method, but only required 3,000 simulations, to find 70 parameter sets with lower error in the metrics of interest than the hand-tuned parameter set. We anticipate further refining the proposed workflow, e.g., incorporating adaptive sampling via Bayesian optimization, can dramatically reduce the number of molecular simulations required to identify parameter sets that accurately predict several physical properties.

\begin{table}[h!]
\caption{Description of abbreviations utilized.\label{tab:ABV}}
\begin{center}
\begin{tabular*}{0.47\textwidth}{l@{\hspace{0.4cm}}l@{\hspace{0.4cm}}}
 \hline
 \hline
 Abbreviation & Expansion \\
 \hline
 AP & Ammonium perchlorate \\
 APE & Absolute percent error \\
 AT & Atom-typing scheme \\
 FIM & Fischer information matrix \\
 GAFF & General AMBER Force Field \\
 GEMC & Gibbs ensemble Monte Carlo \\
 GP & Gaussian process \\
 GPR & Gaussian process regression \\
 HFC & Hydrofluorocarbon \\
 HFC-125 & Pentafluoroethane \\
 HFC-32 & Difluoromethane \\
 LD & Liquid density \\
 LHS & Latin hypercube sampling \\
 LJ & Lennard-Jones \\
 MAPE & Mean absolute percent error \\
 MC & Monte Carlo \\
 MD & Molecular dynamics \\
 RMSE & Root mean square error \\
 SVM & Support vector machine \\
 UCMD & Unit cell mean distance \\
 VLE & Vapor--liquid equilibrium \\
 \hline
 \hline
\end{tabular*}
\end{center}
\end{table}
\clearpage

\section*{Acknowledgements}
R.S.D. and E.J.M. acknowledge funding from the National Science Foundation, Award number OAC-1835630. B.J.B., A.W.D., and E.J.M. acknowledge funding from the National Science Foundation, Award number CBET-1917474. B.J.B. acknowledges funding from the Richard and Peggy Notebaert Premier Fellowship. G.M.T. and E.J.M. acknowledge funding from the Air Force Office of Scientific Research under Contract AFOSR FA9550-18-1-0321. Computational resources were provided by the Notre Dame Center for Research Computing.

\section*{Data and Software Availability}
Codes used to perform the HFC case study and all generated parameters sets are available at: \href{https://github.com/dowlinglab/hfcs-fffit}{https://github.com/dowlinglab/hfcs-fffit}.
Codes used to perform the AP case study and all generated parameter sets are available at: \href{https://github.com/dowlinglab/ap-fffit}{https://github.com/dowlinglab/ap-fffit}.

\bibliography{7_refs.bib}

\end{document}


\title{Supporting Information for: Machine Learning Directed Optimization of Classical Molecular Modeling Force Fields}

\renewcommand{\arraystretch}{0.7}

\author{Bridgette J. Befort}\thanks{BJ Befort and RS DeFever contributed equally to this work.}
\author{Ryan S. DeFever}\thanks{BJ Befort and RS DeFever contributed equally to this work.}
\author{Garrett M. Tow}
\author{Alexander W. Dowling}
\author{Edward J. Maginn}\thanks{Corresponding author}
\affiliation{Department of Chemical and Biomolecular Engineering, University of Notre Dame, Notre Dame, Indiana 46556, United States}

\date{\today}
\maketitle

\renewcommand{\thepage}{S\arabic{page}} 
\renewcommand{\thesection}{S\arabic{section}}  
\renewcommand{\thetable}{S\arabic{table}}  
\renewcommand{\thefigure}{S\arabic{figure}}

\section{Methods}

\subsection*{Identifying parameter sets that are well-separated in parameter space}
The distance between each parameter set is taken as the $L_1$ norm in scaled parameter space. Scaled parameter space is defined such that the lower bound of a parameter is equal to 0.0 and the upper bound is equal to 1.0. The following algorithm was used to select well-separated points: (1) define a distance threshold, (2) select one parameter set at random and add it to the list of those for the next iteration (3) discard all parameter sets within the distance threshold of the parameters sets selected for the next iteration, (4) return to (2) and continue iterating until no parameter sets remain, (5) check the final number of parameter sets identified for the next iteration, and if more than desired, start over and return to (1) with a larger distance threshold.

\section{Discussion}

\subsection*{HFC Identifiability Analysis}
Local identifiability analysis was performed via eigenvalue-eigenvector decomposition of the Fischer information matrix (FIM) which describes sensitivity of the fitted parameters to the experimental data. An FIM with eigenvalues of or near zero is singular and indicates that (some) of the parameters are insensitive to the data (see Chapter 10 of Ref. \citenum{bard1974nonlinear} and Section 3.4 of Ref. \citenum{seber1989nonlinear}). To build the FIM, first a Jacobian matrix $J^T$ was approximated with the central finite difference formula. Properties resulting from the perturbed parameter sets and used in the gradient calculation were obtained from the GP surrogate models or simulations. The product of the Jacobian and its transpose, $J^T \cdot J$, approximates the FIM.  Eigenvalue-eigenvector decomposition was performed on the FIM. The number of non-zero eigenvalues of the FIM indicate the number of directions which are identifiable and the eigenvectors corresponding to near-zero eigenvalues indicate the directions of unidentifiability. Each component in an eigenvector corresponds with a single parameter. If one of these components is of a much larger order of magnitude than the other components in an eigenvector (or the other components were zero), the parameter corresponding to that component would be unidentifiable. However, the components of the eigenvector could all be non-zero and of similar order of magnitude, indicating that the unidentifiability is in a direction that is the linear combination of all of the parameters. 

We applied this analysis to the top four force fields for both HFCs. For the HFC identifiability analyses which used only liquid density, we found there was a single direction of unidentifiability that was a linear combination of parameters for the HFC-32 case and five directions of unidentifiability that were linear combinations of parameters for the HFC-125 case. Upon adding the VLE data into the sensitivity analysis, the models for both HFCs became fully identifiable. We performed the identifiability analysis two different ways: in the first case, we used the GP models to build the Jacobian matrix, and in the second case we performed additional molecular simulations to build the Jacobian matrix. In both cases, we obtained the same conclusions, indicating once again that the GP models are very good at predicting the results from molecular simulations.

Eigenvalue and eigenvector results using GP and simulation predictions for each HFC for liquid density and VLE data are included in the Supporting Information spreadsheets in the zip files `HFC32-Identifiability.zip' and `HFC125-Identifiability.zip'.

\clearpage


\section{Figures and Tables}

\begin{figure}[ht]
\centering
\includegraphics[width=0.5\textwidth]{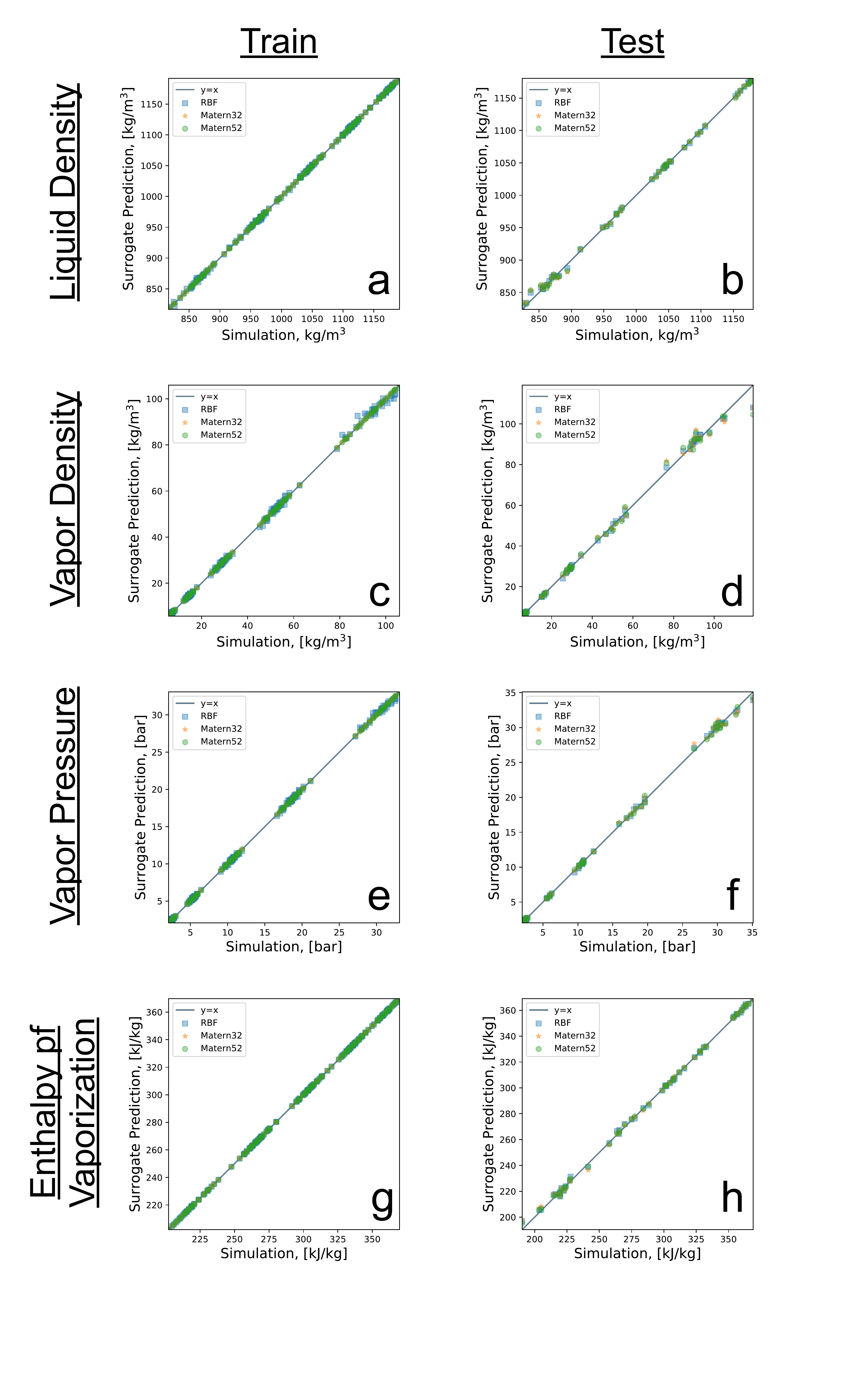}
\caption{The simulation result compared with the GP surrogate model prediction for the surrogate models trained during the VLE-2 iteration for HFC-32. Comparisons are shown for $\rho^l_\mathrm{sat}$ (a, b), $\rho^v_\mathrm{sat}$ (c, d), $P_\mathrm{vap}$ (e, f), $\Delta H_\mathrm{vap}$ (g, h). Comparisons for the training data are shown in the left column (a, c, e, g) and the comparisons for the test data are shown in the right columns (b, d, f, h). Radial basis function (RBF), Mat\'ern $\nu=3/2$ (Matern32), and Mat\'ern $\nu=5/2$ (Matern52) refer to the kernel for the GP surrogate models \cite{rasmussen2003gaussian}. These results are representative of the GP model accuracy for LD and VLE iterations of the HFC force field optimization.}
\label{fig:fig-train-test}
\end{figure}

\begin{figure}[ht]
\centering
\includegraphics[width=0.8\textwidth]{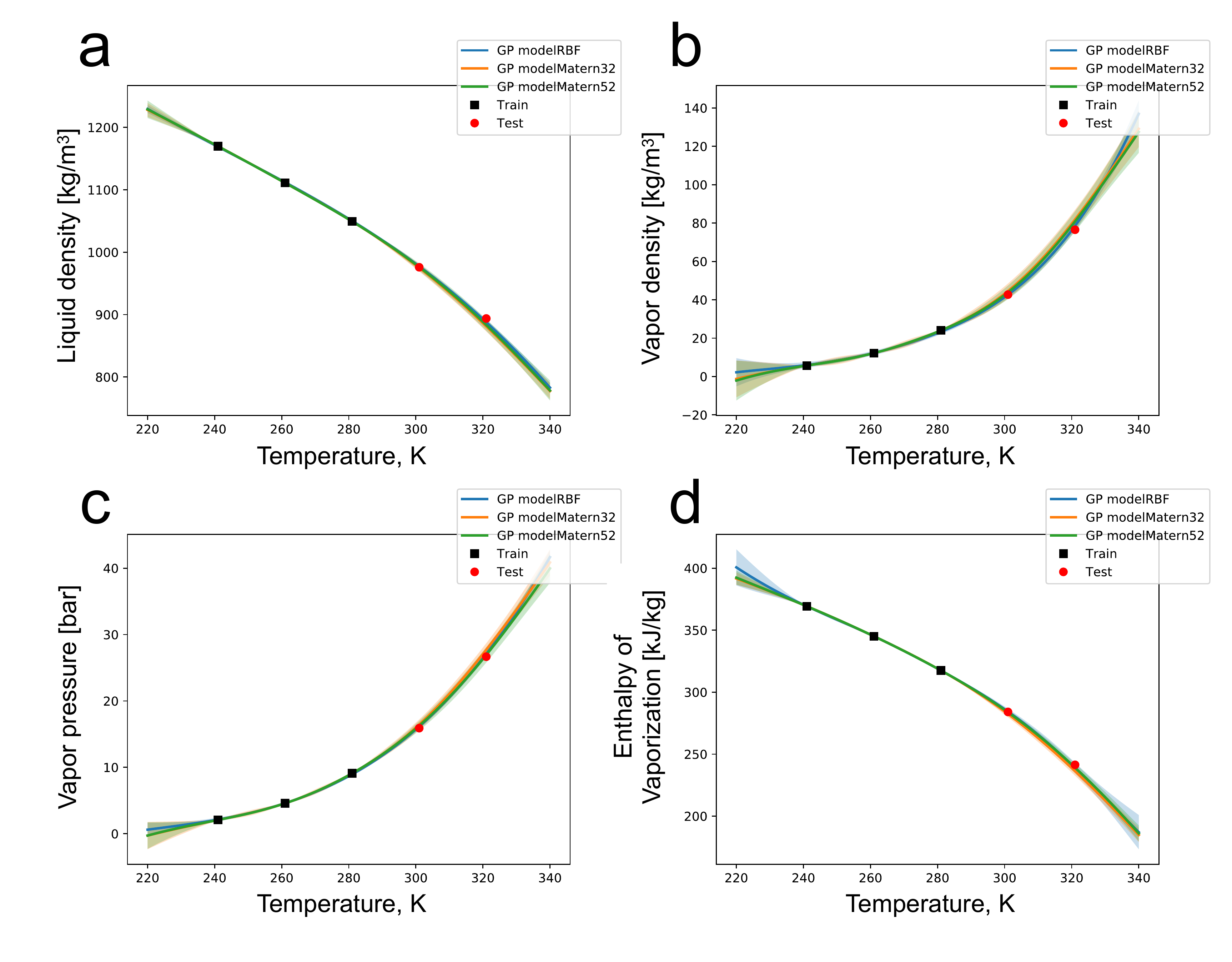}
\caption{Examples of the GP surrogate model means (lines) and variances (shaded regions) for one parameter set from the VLE-2 iteration for HFC-32. Radial basis function (RBF), Mat\'ern $\nu=3/2$ (Matern32), and Mat\'ern $\nu=5/2$ (Matern52) refer to the kernel for the GP surrogate models \cite{rasmussen2003gaussian}. Points shown in black were included in the training data for the GP models, whereas points in red were excluded. GP surrogate models shown for $\rho^l_\mathrm{sat}$ (a), $\rho^v_\mathrm{sat}$ (b), $P_\mathrm{vap}$ (c), $\Delta H_\mathrm{vap}$ (d). These results are representative of the GP model accuracy for LD and VLE iterations of the HFC force field optimization.}
\label{fig:fig-slices}
\end{figure}

\begin{figure}[ht]
\centering
\includegraphics[width=0.8\textwidth]{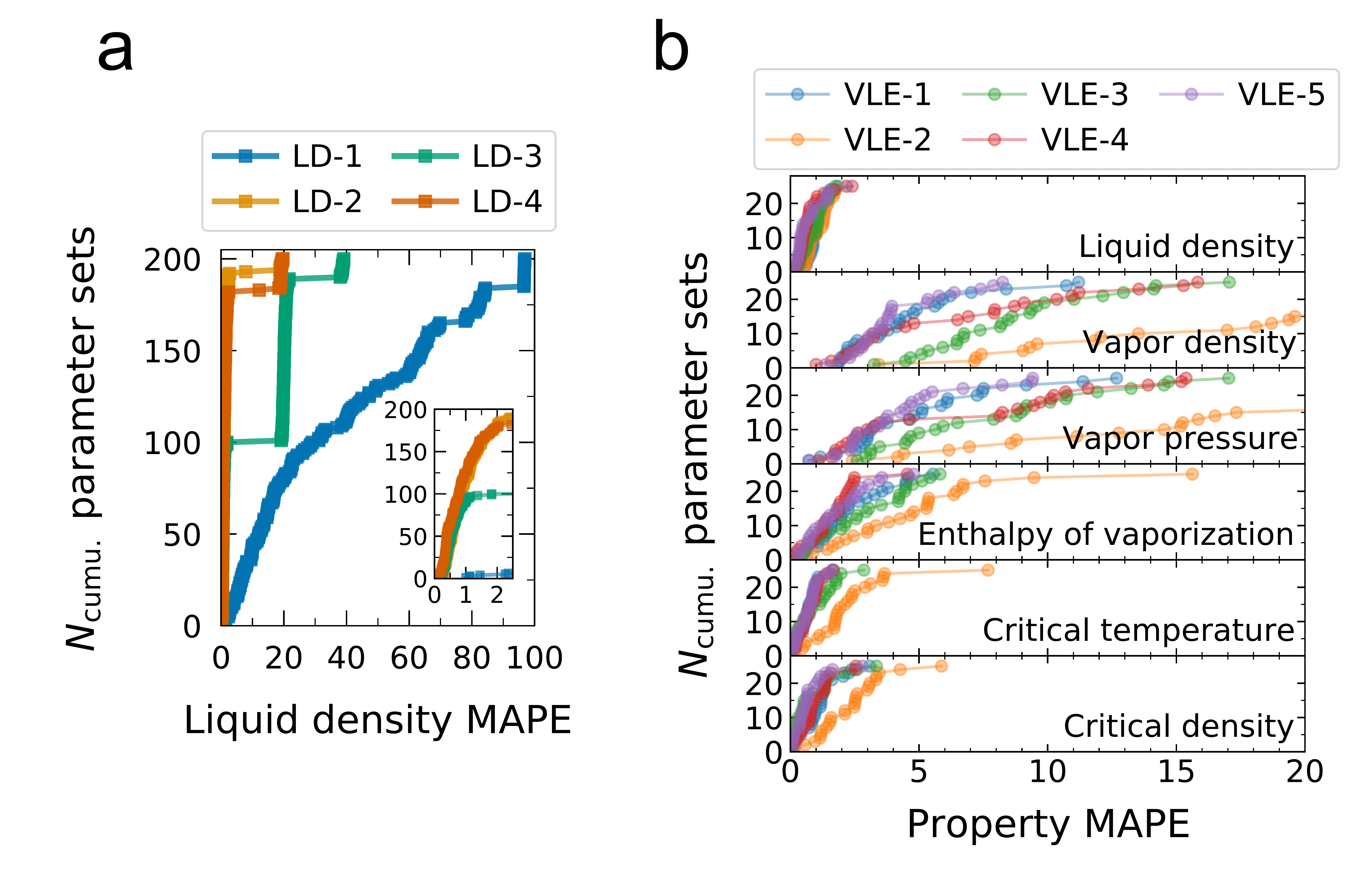}
\caption{Cumulative number of HFC-125 parameter sets per iteration with less than some MAPE for (a) the liquid density iterations 1--4 (LD-$n$) and (b) vapor--liquid equilibrium iterations 1--5 (VLE-$n$), where $n$ is the iteration number. Inset in panel (a) shows the LD behavior for liquid density MAPE $<2.5$\%.}
\label{fig:fig-r125-errors}
\end{figure}

\begin{figure}[ht]
\centering
\includegraphics[width=0.8\textwidth]{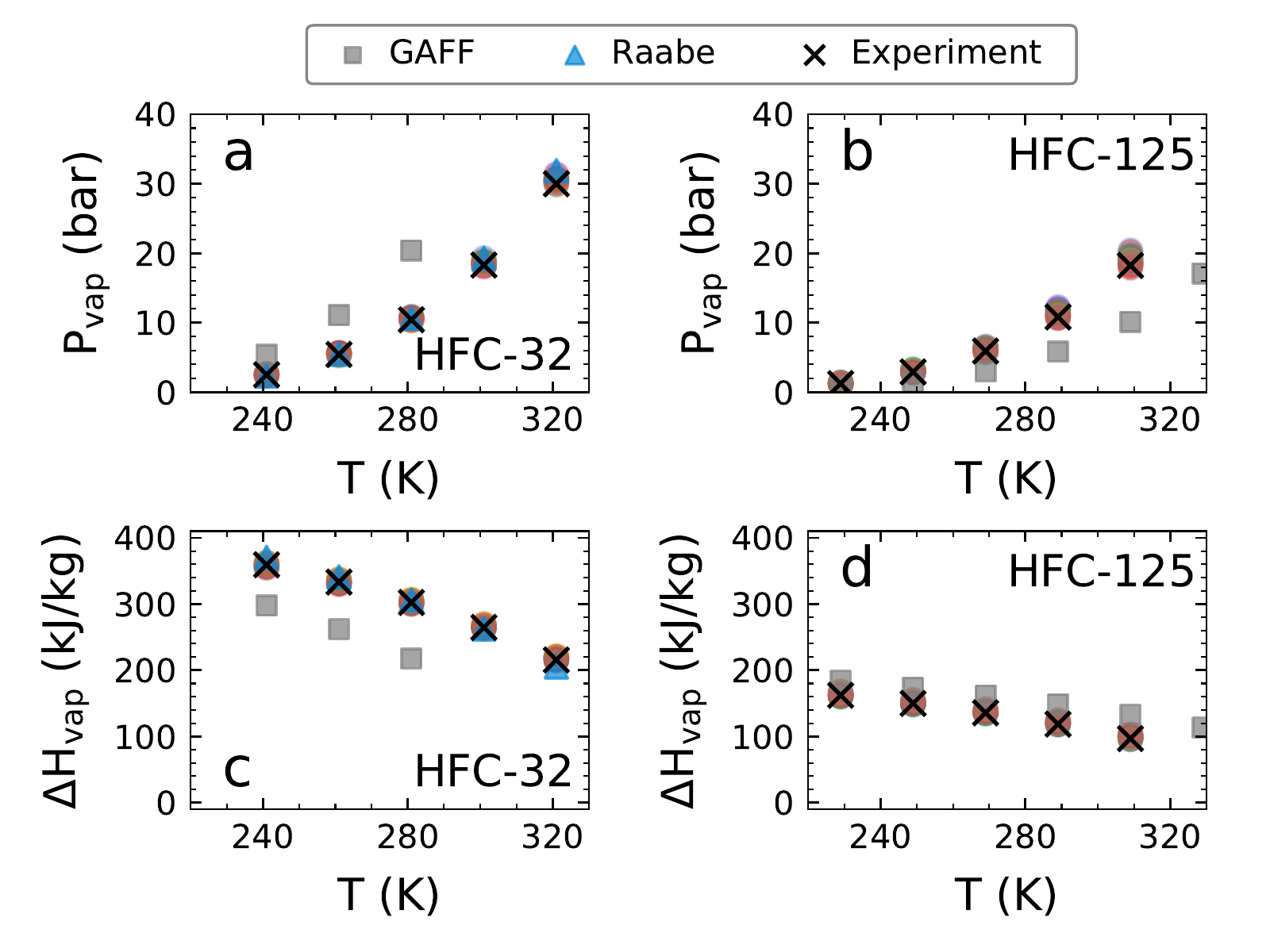}
\caption{Vapor pressure and enthalpy of vaporization for HFC-32 and HFC-125 force fields compared with literature \cite{wang2004development, raabe2013molecular} and experiment \cite{lemmon2018nist}. The 26 (HFC-32) and 45 (HFC-125) non-dominated parameter sets are shown as lightly shaded colored circles. All the non-dominated parameter sets for both HFCs well reproduce the experimental values and are thus highly overlapped.}
\label{fig:fig-enthvap-pvap}
\end{figure}

\begin{figure}[ht]
\centering
\includegraphics[width=0.8\textwidth]{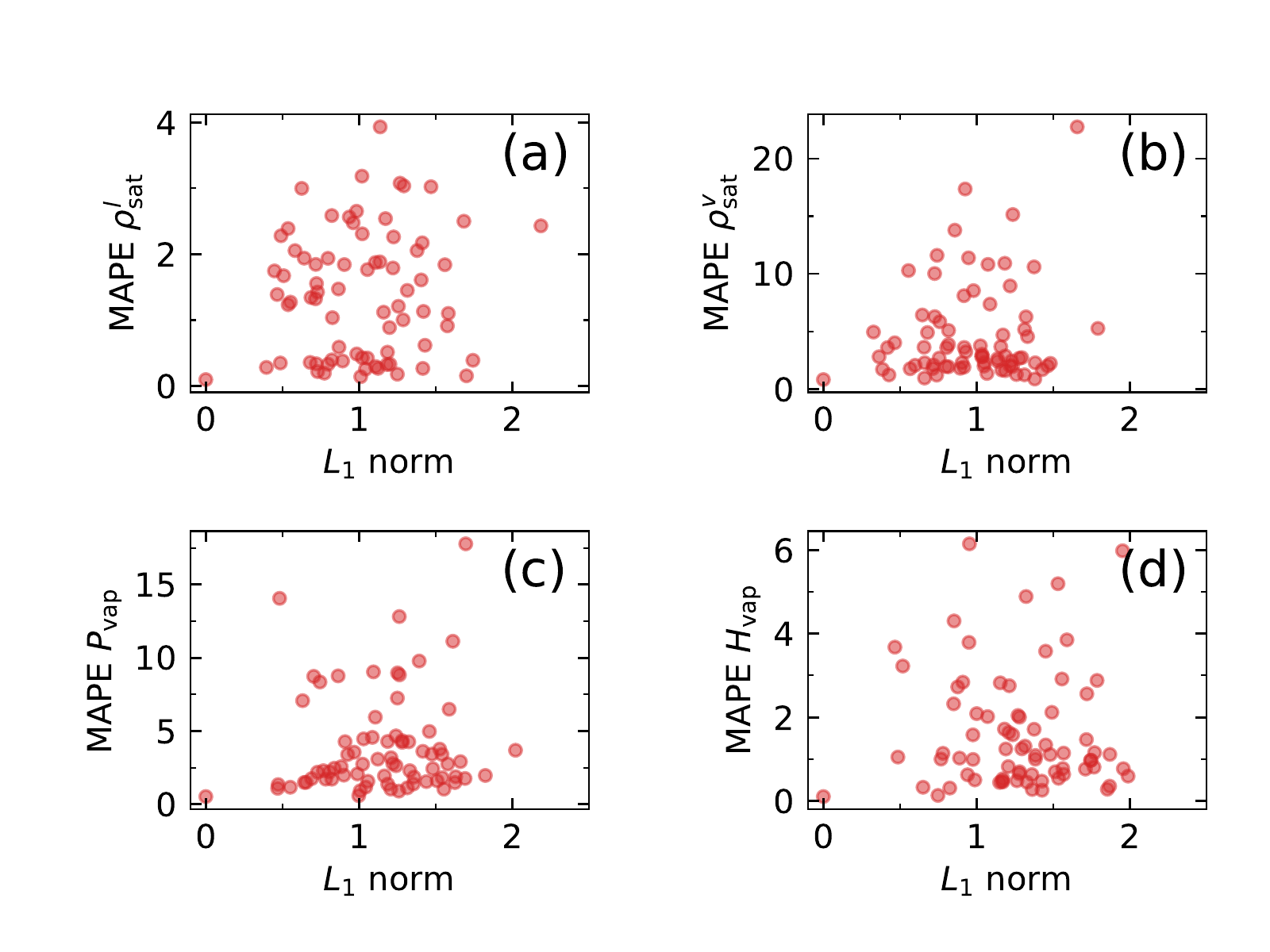}
\caption{Distance between the best parameter set for each property ($L_1$ norm with normalized parameter values) and all other parameter sets versus the property error for HFC-32 VLE iterations. The point with an $L_1$ norm of 0.0 shows the performance of the best parameter set for a given property. One point is shown for each parameter set tested during the VLE iterations. The lack of correlation between the $L_1$ distance from the top performing parameter set and the property error emphasizes that high quality parameter sets are distributed throughout parameter space.}
\label{fig:parameter-comp}
\end{figure}

\clearpage

\begin{table}[ht]
    \centering
    \caption{Partial charges and intramolecular parameters for HFC-32 and HFC-125}
    \label{tab:hfc-intra}
    \begin{tabular}{c c c c}
    \toprule
    \multicolumn{4}{c}{Partial Charges} \vspace{0.2cm} \\
    Type & GAFF Type & $q$ ($e$) & \\
    \colrule
    C  & c3 & 0.405467  &\\
    F  & f  & -0.250783 &\\
    H  & h2 & 0.0480495 &\\
    C1 & c3 & 0.224067  &\\
    C2 & c3 & 0.500886  &\\
    F1 & f  & -0.167131 &\\
    F2 & f  & -0.170758 &\\
    H1 & h2 & 0.121583  &\\
    \colrule
    \multicolumn{4}{c}{Bonds} \vspace{0.2cm} \\
    GAFF Type & $k_r$ (kcal~mol\textsuperscript{-1}~\AA\textsuperscript{-2}) & $r_\mathrm{0}$ (\AA)& \\
    \colrule
    c3-f  & 356.9 & 1.3497 & \\
    c3-h2 & 331.7 & 1.0961 &  \\
    c3-c3 & 300.9 & 1.5375 & \\
    \colrule
    \multicolumn{4}{c}{Angles} \vspace{0.2cm} \\
    GAFF Type & $k_\theta$ (kcal~mol\textsuperscript{-1}~rad\textsuperscript{-2}) & $\theta_\mathrm{0}$ (deg)& \\
    \colrule
    f-c3-f   & 70.9 & 107.36 & \\
    f-c3-h2  & 51.1 & 108.79 & \\
    c3-c3-f  & 66.1 & 109.24 &\\
    c3-c3-h2 & 46.2 & 110.22 & \\
    h2-c3-h2 & 39.0 & 110.20 & \\
    \colrule
    \multicolumn{4}{c}{Dihedrals} \vspace{0.2cm} \\
    GAFF Type & $\nu_n$ (kcal~mol\textsuperscript{-1}) & $n$ & $\gamma$ (deg) \\
    \colrule
    f-c3-c3-f  & 1.20   & 1 & 180.0 \\
    f-c3-c3-h2 & 0.1556 & 3 & 0.0 \\
    \botrule
    \end{tabular}
\end{table}

\begin{table}[ht]
    \centering
    \caption{HFC-32 force field tuning parameters}
    \label{tab:ap-ff}
    \begin{tabular}{c c c c}
    \toprule
    \multicolumn{3}{c}{Intermolecular parameters} \vspace{0.2cm} \\
    Type &  $\sigma$ Bounds (\AA) & $\varepsilon$ Bounds (K/$k_B$) \\
    \colrule
    C &  3.0--4.0 & 20.0--60.0\\
    F &  2.5--3.5 & 15.0--40.0\\
    H &  1.7--2.7 & 2.0--10.0\\

    \botrule
    \end{tabular}
\end{table}

\begin{table}[ht]
    \centering
    \caption{HFC-125 force field tuning parameters}
    \label{tab:ap-ff}
    \begin{tabular}{c c c c}
    \toprule
    \multicolumn{3}{c}{Intermolecular parameters} \vspace{0.2cm} \\
    Type &  $\sigma$ Bounds (\AA) & $\varepsilon$ Bounds (K/$k_B$) \\
    \colrule
    C1 &  3.0--4.0 & 20.0--60.0\\
    C2 &  3.0--4.0 & 20.0--60.0\\
    F1 &  2.5--3.5 & 15.0--40.0\\
    F2 &  2.5--3.5 & 15.0--40.0\\
    H &  1.7--2.7 & 2.0--10.0\\

    \botrule
    \end{tabular}
\end{table}

\begin{table}[ht]
    \centering
    \caption{Hand-tuned AP force field parameters}
    \label{tab:ap-ff}
    \begin{tabular}{c c c c}
    \toprule
    \multicolumn{4}{c}{Intermolecular parameters} \vspace{0.2cm} \\
    Type & $q$ ($e$) & $\sigma$ (\AA) & $\varepsilon$ (kcal/mol) \\
    \colrule
    Cl & 1.5456   & 3.9140 & 0.5018\\
    H  & 0.387625 & 1.7361 & 0.0027\\
    N  & -0.5505  & 3.3078 & 0.0406\\
    O  & -0.6364  & 3.3107 & 0.0954\\
    \colrule
    \multicolumn{4}{c}{Bonds} \vspace{0.2cm}\\
    Type & $k_r$ (kcal~mol\textsuperscript{-1}~\AA\textsuperscript{-2}) & $r_\mathrm{0}$ (\AA) \\
    \colrule
    Cl-O & 426.42 & 1.4523 \\
    H-N  & 413.55 & 1.0300 \\
    \colrule
    \multicolumn{4}{c}{Angles}\vspace{0.2cm}  \\
    Type & $k_\theta$ (kcal~mol\textsuperscript{-1}~rad\textsuperscript{-2}) & $\theta_\mathrm{0}$ (deg) \\
    \colrule
    H-N-H  & 33.45  & 109.5 \\
    O-Cl-O & 107.60 & 109.5 \\
    \botrule
    \end{tabular}
\end{table}

\begin{table}[ht]
    \centering
    \caption{AP force field tuning parameters}
    \label{tab:ap-ff}
    \begin{tabular}{c c c c}
    \toprule
    \multicolumn{3}{c}{Intermolecular parameters} \vspace{0.2cm} \\
    Type &  $\sigma$ Bounds (\AA) & $\varepsilon$ Bounds (kcal/mol) \\
    \colrule
    Cl &  3.5--4.5 & 0.1--0.8\\
    H  &  0.5--2.0 & 0.0--0.02\\
    N  &  2.5--3.8 & 0.01--0.2\\
    O  &  2.5--3.8 & 0.02--0.3\\

    \botrule
    \end{tabular}
\end{table}


\addtolength{\tabcolsep}{1pt}
\begin{table}[ht]
    \centering
    \caption{Screening criteria for AP iterations}
    \label{tab:ap-ff-screening-criteria}
    \begin{tabular}{c c c c c}
    \toprule
    Iteration & Structure Classifier & Symmetry Classifier       & UCMD            & Lattice MAPE \\
              & Threshold (\AA)      &  Threshold  (\AA) & Threshold (\AA) &  Threshold \\
    \colrule
    1-2  & 0.8 & - & 0.35 & 2.5\\
    2-3  & 0.8 & - & 0.35 & 2.5\\
    3-4  & 0.8 & 0.001 & 0.2 & 1.5\\
    \botrule
    \end{tabular}
\end{table}

\begin{table}[ht]
    \centering
    \caption{Critical temperatures ($T_c$) and densities ($\rho_c$) predicted by GAFF \cite{wang2004development}, the force field of Raabe \cite{raabe2013molecular}, and the top four HFC-32 parameter sets compared to experiment \cite{lemmon2018nist}}
    \label{tab:hfc32-cp}
    \begin{tabular}{c c c}
    \toprule
    Force Field & $T_c$ (K) & $\rho_c$ (kg/m$^3$) \\
    \colrule
    GAFF        & 315.3 & 400.1\\
    Raabe       & 344.1 & 430.9\\
    Top A       & 351.1 & 431.0\\
    Top B       & 352.8 & 430.5\\
    Top C       & 351.9 & 431.8\\
    Top D       & 352.9 & 430.9\\
    Experiment  & 351.4 & 429.8\\
    \colrule
    \botrule
    \end{tabular}
\end{table}

\begin{table}[ht]
    \centering
    \caption{Critical temperatures ($T_c$) and densities ($\rho_c$) predicted by GAFF \cite{wang2004development} and top four HFC-125 parameter sets compared to experiment \cite{lemmon2018nist}}
    \label{tab:hfc125-cp}
    \begin{tabular}{c c c}
    \toprule
    Force Field & $T_c$ (K) & $\rho_c$ (kg/m$^3$) \\
    \colrule
    GAFF        & 370.0 & 523.4\\
    Top A       & 342.5 & 570.9\\
    Top B       & 341.5 & 562.9\\
    Top C       & 341.8 & 567.5\\
    Top D       & 343.1 & 576.6\\
    Experiment  & 339.4 & 571.9\\
    \colrule
    \botrule
    \end{tabular}
\end{table}

\addtolength{\tabcolsep}{-1pt}

\begin{table}[ht]
    \centering
    \caption{Performance of HFC-32 and HFC-125 force fields with shared atom types. Results reported for the simulated (sim.) and surrogate model (sur.) predictions. The simulated results for HFC-32 with AT-2 are not reported as the highest temperature GEMC simulation was unstable.}
    \label{tab:my_label}
    \begin{tabular}{c c c c c c c c c}
    \toprule
        & \multicolumn{4}{c}{HFC-32 MAPE} & \multicolumn{4}{c}{HFC-125 MAPE} \\
        & $\rho^l_\mathrm{sat}$ & $\rho^v_\mathrm{sat}$ & $P_\mathrm{vap}$ & $\Delta H_\mathrm{vap}$  & $\rho^l_\mathrm{sat}$ & $\rho^v_\mathrm{sat}$ & $P_\mathrm{vap}$ & $\Delta H_\mathrm{vap}$ \\
        \colrule
        AT-2 (sim.) & - & - & - & - & 1.5 & 31.6 & 27.6 & 15.4 \\
        AT-2 (sur.) & 2.3 & 43.6 & 34.8 & 1.3 & 2.0 & 46.7 & 38.2 & 16.2 \\
        \colrule
        AT-3 (sim.) & 0.8 & 2.4 & 1.8 & 1.8 & 2.8 & 4.5 & 2.4 & 4.8\\
        AT-3 (sur.) & 0.8 & 2.3 & 2.2 & 2.0 & 2.7 & 4.0 & 3.0 & 3.8\\
        \colrule
        AT-4 (sim.) & 1.5 & 2.2 & 1.8 & 1.5 & 0.4 & 0.5 & 2.2 & 1.4 \\
        AT-4 (sur.) & 1.4 & 2.6 & 1.9 & 1.5 & 0.3 & 3.2 & 3.8 & 1.5 \\
        \botrule
    \end{tabular}
\end{table}

\begin{table}[ht]
    \centering
    \caption{MAPE of HFC-125 force fields with reduced number of atom types. Results reported for the simulated (sim.) and surrogate model (sur.) predictions.}
    \label{tab:my_label}
    \begin{tabular}{c c c c c}
    \toprule
        & \multicolumn{4}{c}{HFC-125 MAPE} \\
        Atom types & $\rho^l_\mathrm{sat}$ & $\rho^v_\mathrm{sat}$ & $P_\mathrm{vap}$ & $\Delta H_\mathrm{vap}$  \\
        \colrule
        C1, C2, F, H (sim.) & 0.7 & 3.4 & 3.3 & 2.2\\
        C1, C2, F, H (sur.) & 0.5 & 2.5 & 0.5 & 2.4\\
        \colrule
        C1, C2, F, H (sim.)  & 0.8 & 4.1 & 4.8 & 1.7 \\
        C1, C2, F, H (sur.) & 0.9 & 0.9 & 2.0 & 2.0 \\
        \colrule
        C, F, H (sim.)  & 0.4 & 2.8 & 1.2 & 1.7 \\
        C, F, H (sur.)  & 0.5 & 2.5 & 2.4 & 1.1 \\
        \colrule
        C, F, H (sim.)  & 0.5 & 3.0 & 1.3 & 1.9 \\
        C, F, H (sur.)  & 0.5 & 1.1 & 1.9 & 2.1 \\
        \botrule
    \end{tabular}
\end{table}

\clearpage

\bibliography{9_refs_SI.bib}